\documentclass[]{article}

\pdfoutput=1

\usepackage{amsmath}
\usepackage{graphicx}
\usepackage{natbib}
\usepackage[notcite, notref, final]{showkeys}
\usepackage{listings}
\newcommand{\refeq}{Eq.}
\newcommand{\refEq}{Eq.}
\newcommand{\reffig}{Fig.}
\newcommand{\refsec}{sec.}

\newcommand{\reftab}{Table}
\newcommand{\snr}{\text{SNR}}
\newcommand{\snrobs}{\text{SNR}_{\text{obs}}}
\newcommand{\snrobsperf}{\text{SNR}_{\text{perf. obs}}}
\newcommand{\snrmod}{\text{SNR}_{\text{mod}}}
\newcommand{\rpc}{\text{RPC}}
\newcommand{\rpcperf}{\text{RPC}_{\text{perf}}}
\newcommand{\pcobs}{\text{PC}_{\text{obs}}}
\newcommand{\pcmod}{\text{PC}_{\text{mod}}}

\title{A Bayesian framework for verification and recalibration of ensemble forecasts: How uncertain is NAO predictability?}

\author{Stefan Siegert$^1$\footnote{Corresponding author: Stefan Siegert, Exeter Climate Systems, University of Exeter, North Park Road, Exeter, EX4 4QE, United Kingdom, Email: s.siegert@exeter.ac.uk}, David B. Stephenson$^1$, Philip G. Sansom$^1$,\\Adam A. Scaife$^2$, Rosie Eade$^2$, Alberto Arribas$^2$\\{\small\it $^1$ University of Exeter, Exeter, United Kingdom}\\{\small\it $^2$ Met Office, Hadley Centre, Exeter, United Kingdom}}

\begin{document}

\maketitle

\begin{abstract}
Predictability estimates of ensemble prediction systems are uncertain due to limited numbers of past forecasts and observations.  To account for such uncertainty, this paper proposes a Bayesian inferential framework that provides a simple 6-parameter representation of ensemble forecasting systems and the corresponding observations.  The framework is probabilistic, and thus allows for quantifying uncertainty in predictability measures such as correlation skill and signal-to-noise ratios. It also provides a natural way to produce recalibrated probabilistic predictions from uncalibrated ensembles forecasts.
The framework is used to address important questions concerning the skill of winter hindcasts of the North Atlantic Oscillation for 1992-2011 issued by the Met Office GloSea5 climate prediction system.  Although there is much uncertainty in the correlation between ensemble mean and observations, there is strong evidence of skill: the 95\% credible interval of the correlation coefficient of $[0.19,0.68]$ does not overlap zero. There is also strong evidence that the forecasts are not exchangeable with the observations: With over 99\% certainty, the signal-to-noise ratio of the forecasts is smaller than the signal-to-noise ratio of the observations, which suggests that raw forecasts should not be taken as representative scenarios of the observations. Forecast recalibration is thus required, which can be coherently addressed within the proposed framework.
\end{abstract}

\section{Introduction}

% Climate predictability
Recent studies \citep{riddle2013cfsv2, scaife2014skillful, kang2014prediction} corroborate that state-of-the-art atmosphere-ocean models produce skillful predictions of climate variability on seasonal time scales.
% Introduction - current practice and limitations
The performance of such forecasting systems is generally estimated by calculating summary sample statistics,
such as correlation, over a limited sample of past forecasts and corresponding observations \citep[e.g.,][]{goddard2013verification}.
It is then assumed that future forecasts will exhibit similar performance characteristics \citep{otto2012judging}.

% Why this doesn't help answer some big questions - a distribution-oriented approach is needed
However, such measures-oriented forecast verification \citep{jolliffe2012forecast} provides no inherent information about uncertainty in the reliability and skill of the forecast. 
Uncertainty in forecast quality estimates can be substantial for the small time samples and ensemble sizes typical of climate prediction systems. 
Without proper uncertainty quantification, it is difficult to address important questions for the development and use of climate services such as:
\begin{enumerate}
\item Could the observed skill be due to chance sampling, i.e., natural variability in the observed events and ensemble of forecasts? 
\item How might the skill vary for a different non-overlapping time period (e.g., in the future)?
\item How might the skill vary if a new set of ensemble forecasts were generated over the same hindcast period?
\item Are the forecasts exchangeable with the observations, i.e., do the individual model forecasts have similar properties to the observations?
\item How can non-exchangeable ensemble forecasts be used to create a reliable probability forecast of future observations?
\end{enumerate}
To address such questions it is helpful to propose a statistical model capable of representing the joint distribution of $R$ members of an ensemble forecast, $x_{t,1},\ldots,x_{t,R}$, and their verifying observation $y_t$ over a set of times $t=1,\ldots,N$. 

The importance of an explicit statistical model has been recognized for climate change projections, where
statistical models have been used to formalize assumptions about 
climate model output and the observed present and future climate \citep{tebaldi2005quantifying, sansom2013simple}.
\citet{chandler2013exploiting} argues that a statistical model can make all subjective model assumptions (and limitations) explicit, which leads to transparency in subsequent analyses.
The importance of statistical modeling has also been recognized for weather and seasonal climate forecasting, where the prevailing application is to specify the forecast distribution, i.e., the conditional distribution of the observations, given the raw numerical model output.
Statistical modeling in this context is referred to as forecast recalibration; the goal is to eliminate systematic biases from the numerical model output to improve forecast accuracy.
Commonly used methods for forecast recalibration include Model Output Statistics \citep[MOS, ][]{glahn1972mos}, ensemble dressing \citep{wang2005improvement}, and Non-homogeneous Gaussian Regression \citep[NGR, ][]{gneiting2005calibrated}.
In these recalibration frameworks, the forecasts are not perceived as random quantities, and the full joint distribution of forecasts and observations is not specified.
The present study highlights the benefits of modelling the full joint distribution of forecasts and observations, rather than only the conditional forecast distribution.
The joint distribution captures the variability and dependencies of numerical model forecasts and verifying observations, and thus contains useful information for forecast verification.
The approach of evaluating forecast quality from the joint distribution is known as distributions-oriented verification \citep{murphy1987general}. 
It has not been widely applied because sample sizes of hindcast data sets are usually too small to estimate the joint distribution in sufficient detail.
Parametric modeling has been identified as a useful approach to overcome the curse of dimensionality for distributions-oriented forecast verification \citep[e.g., ][]{murphy1998case, bradley2004distributions}.
In this study we specify the joint distribution of forecasts and observations using a parametric statistical model.
The parameters have to be estimated from a small data set of past forecasts and observations, and are therefore uncertain.
We therefore advocate a framework which uses Bayesian inference to simultaneously estimate the parameters and quantify their uncertainty.
We show how a Bayesian framework can be applied to verification and recalibration of ensemble forecasts based on a small hindcast data set.

So how should one model an ensemble forecasting system so as to capture
the relevant dependencies and variations in forecasts and observations?
In this paper, we study a signal-plus-noise model for an ensemble of runs from a numerical forecast model and the corresponding observations.
The statistical model assumes the existence of a predictable ``signal'' which generates correlation between forecast model runs and observations, as well as the existence of unpredictable ``noise'', which leads to internal variability and random forecast errors.
``Signal'' and ``noise'' are modeled as independent Normally distributed random variables. 
The members of the numerical forecast ensemble are assumed to be exchangeable with one another, i.e. statistically indistinguishable, but not necessarily exchangeable with the observations. 
Possible violations of exchangeability captured by the chosen signal-plus-noise model include a constant bias of the mean, a linear transformation of the predictable signal, as well as differing signal-to-noise ratios.
The signal-plus-noise model is related to the statistical models used by \citet{murphy1990assessment}, \citet{kharin2003improved}, \citet{weigel2009seasonal}, and \citet{kumar2014relationship}. 
We discuss these in more detail in \refsec\ \ref{sec:sn-model}, describe new methods for estimating the model parameters, and present novel applications of the signal-plus-noise model to verification and recalibration of seasonal climate forecasts.

In \refsec\ \ref{sec:application}, the proposed statistical framework is used to analyze recent North Atlantic Oscillation (NAO) hindcasts made with the Met Office GloSea5 seasonal climate prediction system \citep{maclachlan2014global, scaife2014skillful}. 
We demonstrate how the framework allows us to coherently address questions 1-5 above, i.e., to analyze uncertainty in correlation skill, assess the exchangeability of forecasts and observations, and transform raw ensemble forecasts into recalibrated predictive distribution functions.

\section{A signal-plus-noise model for ensemble forecasts}\label{sec:sn-model}

The statistical model used here is motivated by a simple interpretation of ensemble forecasts in the climate sciences, which assumes that observations and forecasts share a common predictable component (the signal), and unpredictable discrepancies arise due to model errors, internal variability, measurement error etc. (the noise).  
Although the same or similar statistical models have been used in previous studies (summarized in \refsec\ \ref{sec:related}), we will provide a detailed discussion of the underlying statistical assumptions and their implications.

\subsection{The signal-plus-noise model}\label{sec:sn-model-sub}

% model equations:
Let $y_t$ be the observation at time $t$, and $x_{t,r}$ the ensemble member (``run'') $r$ at time $t$.
The time $t$ assumes values $1,\cdots,N$, and the ensemble run index $r$ assumes values $1,\cdots,R$.
The model equations are:
\begin{subequations}
\begin{align}
y_t & = \mu_y + \phantom{\beta}s_t + \epsilon_t\\
x_{t,r} & = \mu_x + \beta s_t + \eta_{t,r}\label{eq:model-x}
\end{align}
\label{eq:model}
\end{subequations}
where $\mu_y$, $\mu_x$ and $\beta$ are constants, and $s_t$, $\epsilon_t$, and $\eta_{t,r}$ are assumed to be independent Normal random variables with mean zero and constant variances $\sigma_s^2$, $\sigma_\epsilon^2$, and $\sigma_\eta^2$, respectively.

% interpretation of the random variables
The marginal expected values of the observations $y_t$ and the ensemble members $x_{t,r}$ are equal to $\mu_y$ and $\mu_x$, respectively.
The random variable $s_t$ describes an unobservable ``predictable signal'' shared between forecasts and observations.
The coupling parameter $\beta$ determines the sensitivity of the forecasts to the predictable signal.
The random variable $\epsilon_t$ models the unpredictable component of observed climate, or ``weather noise'', and the random variable $\eta_{t,r}$ models ensemble variability, or ``model noise''.

The model \refeq\ (\ref{eq:model}) includes a number of assumptions about the forecasts and observations.
The data are Normally distributed, and forecasts and observations at different times are conditionally independent, given the model parameters.
Forecasts and observations share a common source of variability, which is modeled by the random variable $s_t$.
The ensemble members are statistically exchangeable with one another, but are generally not exchangeable with the observation.
There exist systematic and/or random discrepancies between model runs and observations, which includes the possibility of a constant model bias ($\mu_x - \mu_y \neq 0$), and possibly different strengths of the predictable signal and unpredictable noise in forecast and observation ($\beta \neq 1$, $\sigma_\epsilon \neq \sigma_\eta$).

% joint normality
We have argued in the Introduction that it is useful to specify a model for the full joint distribution of forecasts and observations. Under the model given by \refeq\ (\ref{eq:model}), forecasts and observations are distributed as a multivariate Normal distribution:
\begin{equation}
(y,x_1, \cdots, x_R)^T \sim \mathcal{N}(\mu, \Sigma)
\end{equation}
with $(R+1)$-dimensional mean vector
\begin{equation}
\mu = (\mu_y, \mu_x, \cdots, \mu_x)^T.
\end{equation}
The $(R+1)\times(R+1)$ dimensional covariance matrix 
$\Sigma$ has entries
\begin{subequations}
\begin{align}
var(y) & = \sigma_s^2 + \sigma_\epsilon^2,\label{eq:vary}\\
var(x_i) & = \beta^2 \sigma_s^2 + \sigma_\eta^2,\label{eq:varxi}\\
cov(x_i, x_j) & = \beta^2 \sigma_s^2\ (i\neq j)\text{, and}\label{eq:covxx}\\
cov(x_i, y) & = \beta \sigma_s^2.\label{eq:covxy}
\end{align}
\label{eq:varcov}
\end{subequations}
for all $i,j = 1,\ldots, R$. Therefore, the model \refeq\ (\ref{eq:model}) can be considered as a simplified parametrization of a covariance matrix of jointly Normal ensemble members and observations, which assumes exchangeability among the ensemble members.
By modeling the $R+1$ observable random variables $y_t$ and $x_{t,r}$ by an unobservable latent variable $s_t$, the number of free parameters in the covariance matrix $\Sigma$ is reduced from $(R+1)(R+2)/2$ to only $4$.
Invoking a latent variable provides a parsimonious description of the joint distribution of forecasts and observations.
Note further that the variance of the ensemble mean is given by
\begin{equation}
var(\bar{x}) = \beta^2\sigma_s^2 + \frac{1}{R}\sigma_\eta^2,\label{eq:varxbar}
\end{equation}
and that the covariance $cov(x_i, y)$ between observations and individual ensemble members is equal to the covariance $cov(\bar{x},y)$ between observations and the ensemble mean.
The correlation skill of the ensemble mean can thus be expressed in terms of the model parameters by
\begin{equation}
\rho = \frac{cov(\bar{x},y)}{\sqrt{var(\bar{x})var(y)}}= \frac{\beta\sigma_s^2}{\sqrt{(\beta^2\sigma_s^2+\sigma_\eta^2/R)(\sigma_s^2 + \sigma_\epsilon^2)}}.\label{eq:rho-pop}
\end{equation}

% interpretation of model parameters
The model parameters can be used to assess further aspects of the quality of the forecasting system.
The forecasts are exchangeable with the observations if and only if $\mu_x=\mu_y$, $\beta=1$, and $\sigma_\epsilon=\sigma_\eta$.
If these conditions are met, the ensemble forecast is perfectly reliable, i.e., the observation is indistinguishable from the ensemble members, and the individual ensemble members can be taken as representative scenarios for the observation.
If the forecast is reliable in the above parametric sense, the additional criterion $\sigma_\epsilon = \sigma_\eta = 0$ indicates a perfect deterministic forecast; all ensemble members are then always exactly equal to the observation.
If, on the other hand, either $\beta=0$ or $\sigma_s=0$, there is no systematic relation between the forecasts and observations, i.e., the forecasts have no skill.
The forecasts are marginally calibrated, i.e., forecast and observed climatology are equal, if $\mu_x = \mu_y$ and $\beta^2\sigma_s^2 + \sigma_\eta^2 = \sigma_s^2 + \sigma_\epsilon^2$.

% interpretation of the signal
The variable $s_t$, referred to as ``the predictable signal'', requires careful interpretation. 
Essentially, this latent variable is a model construct that provides covariance - it cannot be directly observed.
However, for climate predictions the concepts of ``signal'' and ``noise'' can be (and have been) given a physical interpretation (e.g., \citet{madden1976estimates}, \citet[][sec. 17.2.2]{storch2001statistical} or \citet{eade2014seasonal}).
The predictable signal can be understood as the slowly varying component of weather related to longer time scale processes (e.g., ocean circulation). 
The noise is interpreted as weather variability, which cannot be predicted deterministically on time scales of more than a few days.
It should be noted that the signal estimated here is a property of the observations and the forecasts, and it is not a unique property of the ``real world''.
Different forecasting models for the same observation can give rise to different ``signals''.

\subsection{Related statistical models}\label{sec:related}

Related models have been widely used for statistical data analysis, for example,  in structural equation modeling \citep{pearl2000causality}, factor analysis \citep{everitt1984introduction}, latent variable modeling \citep{bartholomew2011latent}, and measurement error models, also known as error-in-variables models \citep{fuller1987measurement, buonaccorsi2010measurement}.
The same or similar models as our signal-plus-noise model \refeq\ (\ref{eq:model}) have also been used to investigate seasonal to decadal climate predictability.
\citet{kharin2003improved} apply the signal-plus-noise model to seasonal climate forecast variability.
Like the present study, \citet{kharin2003improved} use the model to study the relationship between variability and predictability, and also use the explicit statistical assumptions to calibrate imperfect ensemble forecasts to improve probabilistic forecast skill.
Their parameter estimation is essentially based on the method of moments, and parameter uncertainty is not quantified.
The present study extends \citet{kharin2003improved} by carefully quantifying uncertainty in the statistical model parameters as well as all derived quantities, and by incorporating this uncertainty in distributions-oriented forecast verification and forecast recalibration. 
More recently, \citet{kumar2014relationship} used the signal-plus-noise model to study the relationship between perfect skill and actual skill in seasonal ensemble forecasts.
They show that perfect skill, i.e., the ability of the ensemble to predict its own realizations can be lower than actual skill, i.e., the ability of the ensemble to predict the real system. 
We will address actual and perfect model predictability in \refsec\ \ref{sec:sn-ana}, where we study signal-to-noise ratios in forecasts and observations.
Unlike \citet{kumar2014relationship}, the present study quantifies uncertainty in the signal-to-noise ratios.

The proposed signal-plus-noise model also relates to previous
frameworks used to interpret ensembles of climate projections (see \citet{stephenson2012statistical} and references therein). 
\citet{rougier2013second} apply a latent variable models to infer future climate from a collection of exchangeable climate model runs.
\citet{chandler2013exploiting} provides a statistical framework for multi-model ensembles, where runs from one climate model are non-exchangeable with runs from different climate models, and non-exchangeable with the observations.
A related Bayesian framework is used by \citet{tebaldi2005quantifying}, who assume different values of model parameters for present and future climate.  
\citet{annan2010reliability} work under the assumption that ensemble forecasts and observations are fully statistically exchangeable; their model is thus a special case of the signal-plus-noise model with $\beta=1$, $\mu_x=\mu_y$, and $\sigma_\epsilon = \sigma_\eta$.

A noteworthy modification was studied by \citet{weigel2009seasonal}.
The observation is similarly decomposed into signal plus noise, but the ensemble members are modeled by adding a common random error term $d_t$ as well as individual error terms $\eta_{t,r}$ to the predictable signal variable: 
\begin{subequations}
\begin{align}
y_t & = s_t + \epsilon_t\\
x_{t,r} & = s_t + d_t + \eta_{t,r}.
\end{align}
\label{eq:weigel-model}
\end{subequations}
We note that this additive model implies that the covariance between ensemble members is $cov(x_i, x_j) = \sigma_s^2 + \sigma_d^2$, and that the covariance between ensemble members and observations is $cov(x_i, y) = \sigma_s^2$, which implies that 
$cov(x_i, y)$ can never be negative, and $cov(x_i, x_j)$ can never be smaller than $cov(x_i, y)$.
Both scenarios are, however, conceivable in real systems and should at least be allowed by a statistical model.
\refEq\ (\ref{eq:varcov}) shows that model \refeq\ (\ref{eq:model}) does not impose these two restrictions; the only similar restriction is that, according to \refeq\ (\ref{eq:covxx}), $cov(x_i, x_j)$ is always positive.

\subsection{Parameter estimation}\label{sec:estimation}

It is possible to calculate point estimates of the model parameters using the
method of moments. This makes use of the first and second sample moments of the
data and equates them with the corresponding expected values in
\refeq\ (\ref{eq:varcov}).  The estimating equations are given in appendix C. Such moment estimators are discussed by
\citet{moran1971estimating} (in the context of linear structural relationships)
who notes that, if $\sigma_\eta^2$ were known exactly, then the
moment estimators are also the maximum-likelihood estimators, and
that complications can arise due to negative variance estimates which require
modifications of the estimator equations.
Point estimates obtained by method of moments or maximum likelihood estimation
are prone to sampling uncertainty, especially for the small
sample sizes typical of climate prediction systems.  It is therefore important
to quantify uncertainty in the model parameters, using either resampling
methods such as the bootstrap \citep{efron1994introduction}, by frequentist
variance estimators or confidence intervals
\citep[e.g.,][]{fuller1987measurement}, or by Bayesian estimation which we use
here. 

In Bayesian statistics degrees of certainty and uncertainty are expressed by
conditional probabilities, and probabilities are manipulated based on the principle of
coherence, i.e. by using only the addition and multiplication rule of
probability calculus \citep{jaynes2003probability,gelman2004bayesian,lindley2006understanding,robert2007bayesian}. For the present study,
the main object of interest for Bayesian inference is therefore the joint conditional probability distribution over
all unknown quantities (i.e., the model parameters), conditional on all known
quantities (i.e., the hindcast data and observations).  From this
posterior distribution we can derive point estimators, for example, the
posterior mean or mode, and uncertainty intervals, for example the $95\%$
parameter values with highest posterior density.  Denote by $\theta=\{\mu_x,
\mu_y, \beta, \sigma_s, \sigma_\epsilon, \sigma_\eta\}$ the collection of unknown
parameters of the signal-plus-noise model, by $s=\{s_1, \cdots, s_N\}$ the unknown values of the latent
signal variable, and by $\{x,y\}=\{x_{t,1},\cdots,x_{t,R},y_t\}_{t=1}^N$ the
collection of known forecasts and observations from a hindcast experiment.  The
desired posterior distribution for Bayesian estimation is thus $p(\theta, s |
x,y)$.  Derivation of the posterior distribution requires the specification of
a prior probability distribution $\pi(\theta, s)$ over the unknown quantities, which
factors into $\pi(\theta)\prod_{t=1}^N\pi(s_t|\sigma_s)$ in our model. The
prior distribution can be used to incorporate a-priori information 
about the modeled data into the inference process (we discuss the prior distribution for our analysis in \refsec\ \ref{sec:prior-spec}).  Furthermore, the likelihood function is required, which is the probability of the data, given specified values of the model parameters. The likelihood function, denoted by $\ell(x,y|\theta,s)$ can be
calculated from \refeq\ (\ref{eq:model}) using the distribution law of the
Normal distribution, and the independence assumption:
\begin{align} & \ell(x,y|\theta, s)  = \prod_{t=1}^N\left\{p(y_t | \theta, s_t)
\prod_{r=1}^R p(x_{t,r}|\theta, s_t) \right\}\nonumber\\ 
& = \left(2\pi\sigma_\epsilon^2\right)^{-N/2} \left(2\pi\sigma_\eta^2\right)^{-NR/2} \exp\left\{-\frac{1}{2} \sum_{t=1}^N\left[ \left(\frac{y_t - (\mu_y + s_t)}{\sigma_\epsilon}\right)^2 + \sum_{r=1}^R\left(\frac{x_{t,r}-(\mu_x + \beta s_t)}{\sigma_\eta}\right)^2\right]\right\}.\label{eq:likelihood-fun} 
\end{align}
Using the likelihood function and the prior distribution, the 
posterior distribution is then formally calculated by Bayes rule
\begin{equation} 
p(\theta,s | x,y) \propto \ell(x,y | \theta,s) \pi(\theta,s)
\label{eq:posterior}
\end{equation}
where the proportionality constant is independent of $\theta$ and $s$, and depends only on the data.

A closed form expression for the joint posterior distribution using arbitrary prior distributions is not available.
 For this paper, we have thus approximated a fully Bayesian analysis by Markov-chain Monte-Carlo (MCMC) integration \citep{brooks2011handbook}, using the freely available software STAN \citep{stan2014}, interfaced via the R package {\tt rstan} \citep{rstan2014}.
MCMC is an efficient computational technique to simulate random draws from an arbitrary (possibly unnormalized) probability distribution, such as our posterior distribution given by \refeq\ (\ref{eq:posterior}). A MCMC program can thus be regarded as a random number generator that samples from the posterior distribution. Using an appropriate MCMC sampler, we can approximate posterior distributions by smoothed histograms, and posterior expectations by empirical averages of samples drawn from the posterior distribution. The STAN software provides a scripting language to translate a user-specified generative model for the data (such as our signal-plus-noise model \refeq\ \ref{eq:model}) into a MCMC sampler.
 The STAN model code for our analyses is provided in appendix A.
The code shows that the derivation of the likelihood function \refeq\ (\ref{eq:likelihood-fun}) is not really required to implement the MCMC sampler in STAN -- specification of the generative model \refeq\ (\ref{eq:model}) is enough. 
 We have used the No-U-Turn sampler of STAN with all its default settings.
 All our posterior distributions are based on $10^5$ Monte-Carlo samples. 
 These were generated by simulating $8$ parallel Markov chains, each for $10^6$ iterations, after discarding a warm-up period of $10^4$ iterations for initialization of the algorithm.
 The 8 chains were thinned by retaining only every 80th sample to eliminate autocorrelation.
 Our procedure for generating the posterior samples takes about 20 minutes on a desktop computer with $8$ CPUs.
 Reasonable results can, however, be obtained without thinning of the Markov chain, which reduces the time to generate $10^5$ samples to a few seconds.
Potential scale reduction factors close to one \citep{gelman1992inference} and visual inspection of trace plots were taken as evidence for successful convergence and proper mixing of the Markov Chains.

\subsection{Relation between ensemble mean and observations}\label{sec:linreg-corr}

The signal-plus-noise model can be used to learn about the relationship between the observations and the means of the ensemble forecasts.  
It follows from standard Normal theory \citep[e.g.,][sec. 3.2]{mardia1979multivariate} that if the model parameters $\theta$ are known, the conditional distribution of the observation $y_t$ given the ensemble mean $\bar{x}_t$ is 
\begin{equation}
(y_t|\bar{x}_t, \theta) \sim \mathcal{N}\left(\mu_y + \frac{\beta\sigma_s^2}{\beta^2\sigma_s^2+\sigma_\eta^2/R}(\bar{x}_t - \mu_x), \sigma_\epsilon^2 + \sigma_s^2\left(\frac{\sigma_\eta^2}{R\beta^2\sigma_s^2+\sigma_\eta^2}\right)\right).\label{eq:linreg-post}
\end{equation}
In other words, the relationship between the observations $y_t$ and ensemble means $\bar{x}_t$ is described by a simple linear regression model whose intercept, slope and residual variance parameters are functions of the known parameters of the signal-plus-noise model. 
So if there was no uncertainty in the signal-plus-noise parameters, one could use \refeq\ (\ref{eq:linreg-post}) as a basis for post-processing the ensemble means to predict the observations. 
Correcting dynamical forecasts by linear regression, also known as model output statistics \citep[MOS,][]{glahn1972mos}, forms the basis for commonly used post-processing techniques in seasonal forecasting \citep[e.g., ][]{feddersen1999reduction}. 
In \refsec\ \ref{sec:cali-pred}, we will compare the simple linear regression approach with a fully Bayesian posterior predictive approach which accounts for parameter uncertainty.

\citet{eade2014seasonal} use the relation between signal-plus-noise interpretation and linear regression in their post-processing technique for the ensemble mean, and then adjust the distribution of the ensemble members around the post-processed ensemble mean to have the signal-to-noise ratio implied from the correlation, while retaining year-to-year variability in the ensemble spread.
That is, while \refeq\ (\ref{eq:linreg-post}) assumes a constant variance, the method of \citet{eade2014seasonal} allows for time varying ensemble variance.
But \citet{tippett2007estimation} have shown for seasonal precipitation forecasts, that retaining the year-to-year variability of the ensemble variance does not improve the forecasts.
The question of whether the ensemble spread should influence the width of the forecast distribution in seasonal NAO forecasting is not addressed further in this paper.

\section{Application to seasonal NAO hindcasts}\label{sec:application}

\subsection{The data}\label{sec:data}

\begin{figure}
\includegraphics{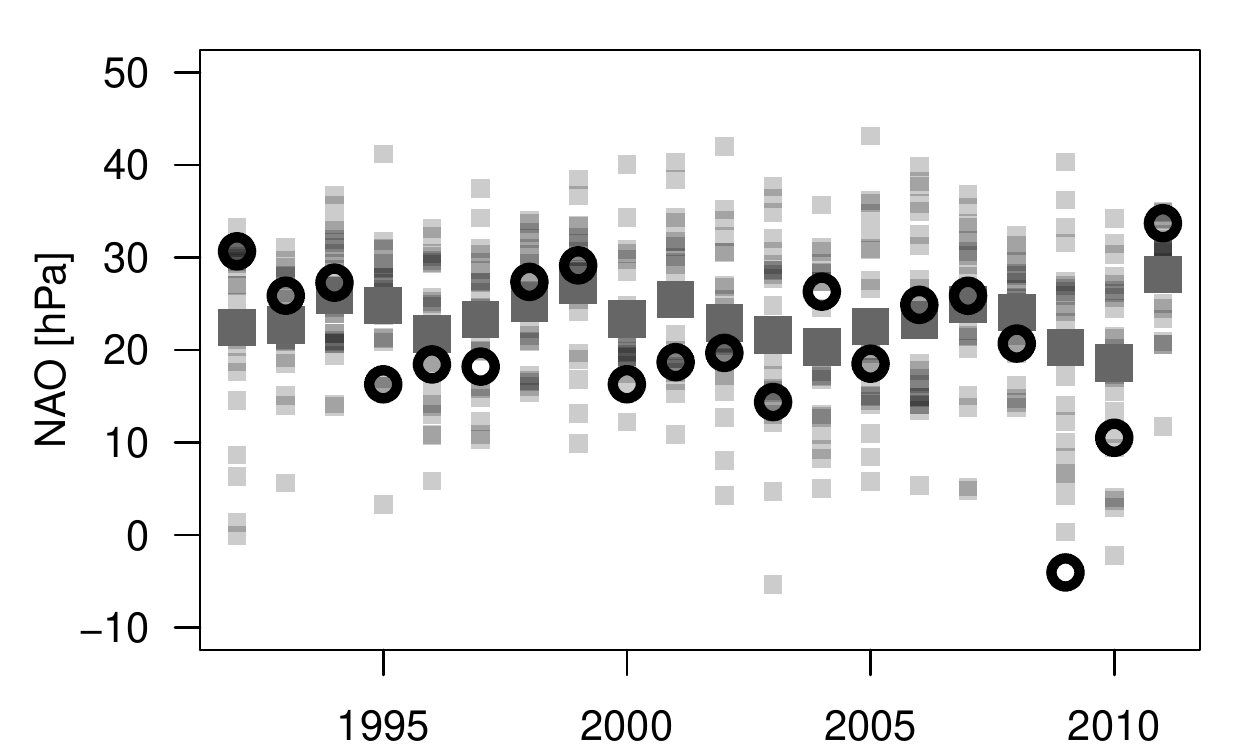}
\caption{Raw winter NAO ensemble data generated by GloSea5 (semi-transparent squares), ensemble mean forecasts (large gray squares), and verifying NAO observations (circles).}\label{fig:nao}
\end{figure}

The signal-plus-noise model is demonstrated here by application to seasonal forecasts 
of the winter (Dec-Feb mean) North-Atlantic oscillation (NAO), discussed in \citet{scaife2014skillful}.
Seasonal NAO predictability has further been studied by \citet{doblas2003skill}, \citet{eade2014seasonal}, and \citet{smith2014seasonal}.
NAO is defined here as the difference in sea-level pressure between the Azores and Iceland (or nearest
model grid points to these two locations; cf. \citet{scaife2014skillful}). 
 A 24-member ensemble hindcast was generated annually from 1992 to 2011 by the Met Office Global Seasonal forecast System 5 (GloSea5), using lagged initialization between 25 October and 9 November (details about GloSea5 can be found in \citet{maclachlan2014global}).
Raw forecast and observation data are shown in \reffig\ \ref{fig:nao}.
In \reftab\ \ref{tab:data-summary} we show a number of summary statistics of the hindcast data.

\begin{table}[hbtp]
\caption{Summary statistics of ensemble means $\bar{x}_t$ and observations $y_t$, and their particular values for the NAO hindcast.}\label{tab:data-summary}
\begin{center}
\begin{tabular}{ccl|cl}
$m_x$ & $=$ & $N^{-1}\sum_{t=1}^N \bar{x}_t$ & $23.42$ & $hPa$ \\[1ex]
$m_y$ & $=$ & $N^{-1} \sum_{t=1}^N y_t$ & $20.94$ & $hPa$ \\[1ex]
$v_{\bar{x}}$ & $=$ & $N^{-1} \sum_{t=1}^N \left(\bar{x}_t - m_x\right)^2$ & $5.24$ & $(hPa)^2$ \\[1ex]
$v_y$ & $=$ & $N^{-1} \sum_{t=1}^N \left(y_t - m_y\right)^2$ & $67.12$ & $(hPa)^2$ \\[1ex]
$s_{\bar{x}y}$ & $=$ & $N^{-1} \sum_{t=1}^N \left(\bar{x}_t - m_x\right)\left(y_t - m_y\right)$ & $11.55$ & $(hPa)^2$\\[1ex]
\end{tabular}
\end{center}
\end{table}

\subsection{Prior specification}\label{sec:prior-spec}

We use the following independent prior distribution functions for the model parameters:
$\mu_x, \mu_y \sim \mathcal{N}(0, 30^2)$, $\sigma_s^2 \sim \mathcal{G}^{-1}(2,25)$, $\sigma_\epsilon^2, \sigma_\eta^2 \sim \mathcal{G}^{-1}(3,100)$, and $\beta\sim \mathcal{N}(1,0.7^2)$, where $\mathcal{G}^{-1}(a,b)$ denotes the Inverse-Gamma distribution with shape parameter $a$ and scale parameter $b$.
A random variable $X \sim \mathcal{G}^{-1}(a,b)$ has a density function proportional to $x^{-a-1}\exp(-b/x)$. 
The Inverse-Gamma distribution was chosen as a prior because it is a common choice for variance parameters which can simplify Bayesian calculations.
 The prior distributions on $\mu_x$ and $\mu_y$ are very wide and uninformative, and we found the inference to be insensitive to the choice of these prior distributions. 
 We found that the inference is more sensitive to the choice of priors on $\beta$ and the $\sigma$ parameters.
These priors were deliberately chosen to be rather narrow:
It can be shown by simulation experiments that, under the prior distributions above, $\sigma_s$ has prior mean $\approx 4hPa$ and prior standard deviation of $\approx 2hPa$, and $\sigma_\eta$ and $\sigma_\epsilon$ both have prior mean of $\approx 6.5hPa$ and prior standard deviation of $\approx 2.5hPa$.
The parameters of the prior distributions were chosen by trial and error to yield reasonable prior distributions on observable quantities.
In particular, the prior distributions of the standard deviation of the ensemble members ($\sqrt{var(x_i)}$, cf. \refeq\ \ref{eq:varxi}), and of the observation ($\sqrt{var(y)}$, cf. \refeq\ \ref{eq:vary}) both have prior mean of $\approx 8hPa$ and prior standard deviation of $\approx 3hPa$.
The correlation coefficient of the 24-member ensemble mean and the observations (cf. \refeq\ \ref{eq:rho-pop}) has prior mean of $\approx 0.4$ and prior standard deviation of $\approx 0.3$, which covers sample correlation coefficients observed in past studies of seasonal winter NAO predictability (see, e.g., \citet{kang2014prediction} and \citet{shi2015impact} for collections of seasonal winter NAO correlations obtained by different models).
Furthermore, the prior probability of the model having lower signal-to-noise ratio than the observation is $\approx 0.5$.
The prior distributions on the model parameters therefore provide reasonable prior specifications for the analyses of \refsec\ \ref{sec:unc-corr} (correlation coefficients) and \refsec\ \ref{sec:sn-ana} (signal-to-noise ratios).
It is worthwhile to point out that the priors are for horizontal atmospheric pressure differences measured in $hPa$; if NAO were measured differently, the above prior distributions would have to be rescaled.

The prior distribution is a subjective choice in Bayesian analysis, and is therefore often subject to criticism and discussion.
We thus want to describe in more detail how we have arrived at the above distributions, and why we found default ``uninformative'' distributions unsatisfactory.
We had initially specified independent uniform prior distributions on the model parameters as follows: $\sigma_{s,\epsilon,\eta} \sim U(0,30)$, and $\beta \sim U(-1,2)$ to cover physically meaningful ranges of the parameter values, but without favoring a priori one set of parameters over the other.
We have sampled model parameters from these prior distributions, and substituted the samples into the analytic expressions of the correlation coefficient given by \refeq\ (\ref{eq:rho-pop}).
A smoothed histogram of the thus transformed samples approximates the derived prior distribution of the correlation coefficient.
We found that the derived correlation prior has multiple modes, two of which are close to $+1$ and $-1$. Since the prior distribution should encode a priori information (for example about NAO prediction skill), this distribution is clearly unjustified.
This example shows how seemingly ``objective'' and ``uninformative'' uniform prior distributions for the model parameters can lead to very informative and physically unjustified prior distributions on meaningful observable quantities.
The uniform priors further produced a prior probability of over $0.6$ that the model has a lower signal-to-noise ratio (SNR) than the observation. 
Since the possible anomalous signal-to-noise ratio in NAO predictions is a questions which we wanted to address, we did not want to bias the result a priori into the direction of a low model SNR.
The chosen prior distributions represent a compromise between subjective judgements and previously published results about NAO variability, signal-to-noise ratio, and correlation skill.

In appendix D, the sensitivity to varying prior specification is illustrated for the correlation analysis of \refsec\ \ref{sec:unc-corr}.
The analysis shows that while different priors indeed lead to different posteriors, the updated posterior distributions are similar.
For more detailed discussions about the role and specification of prior distributions the reader is referred to the standard texts on Bayesian statistics given in \refsec\ \ref{sec:estimation}, in particular \citet{gelman2004bayesian}.
We lastly note that if sufficient data is available, the influence of the prior disappears and the Bayesian inference is dominated by the likelihood function \citep{gelman2013not}.

\subsection{Bayesian updating}\label{sec:bayes-updating}

\begin{figure}
\centering\includegraphics{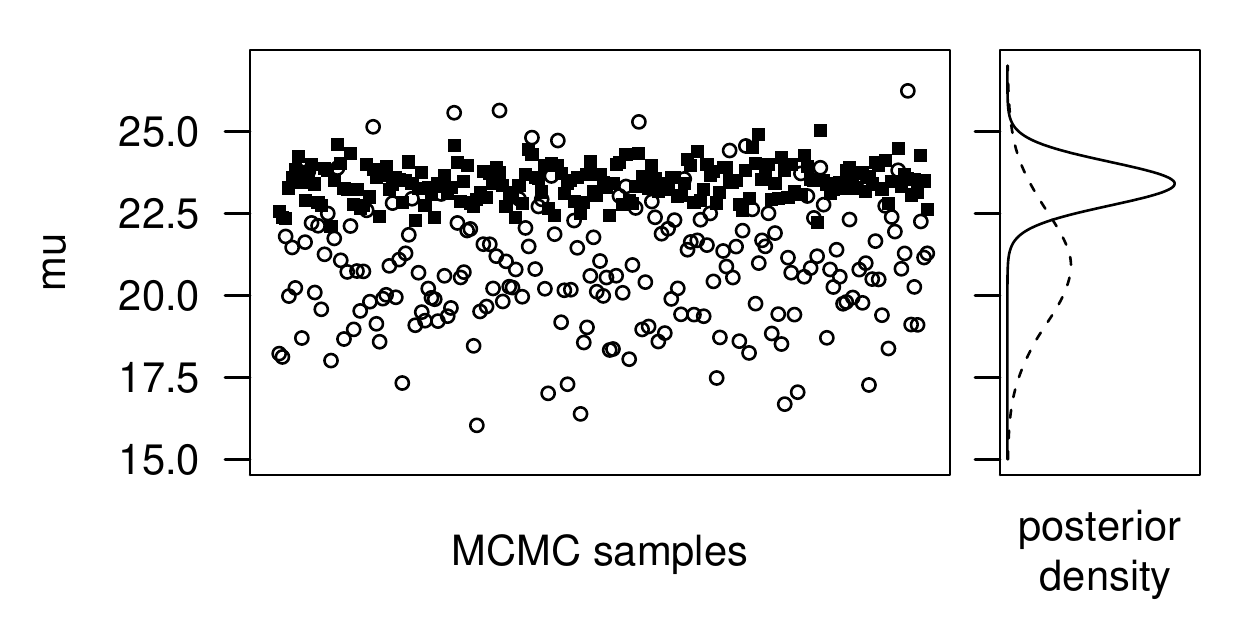}
\caption{Illustration of MCMC approximation of posterior distributions. Left panel: Trace plot of 200 joint samples from the Markov Chain of $\mu_x$ (squares) and $\mu_y$ (circles). Right panel: Posterior distributions of $\mu_x$ (full line) and $\mu_y$ (dashed line), reconstructed from all $10^5$ samples.}\label{fig:mu-chain}
\end{figure}

\begin{figure}
\centering\includegraphics{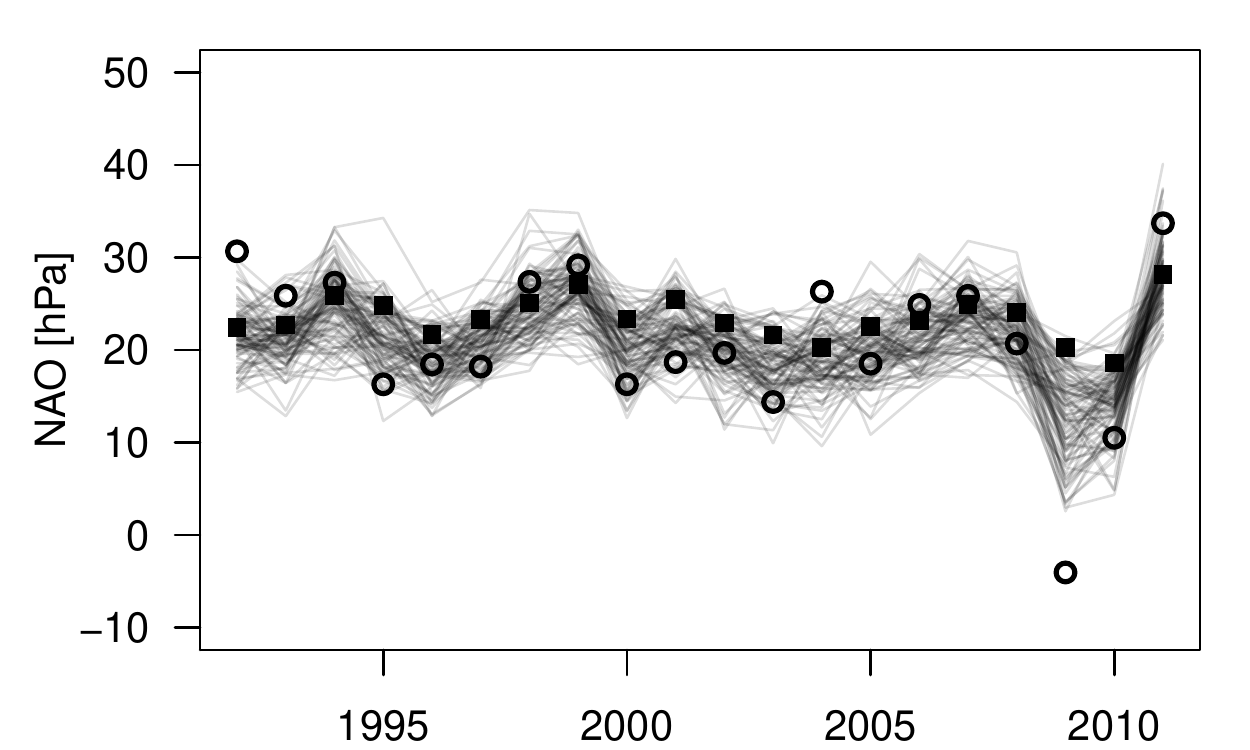}
\caption{NAO observations (circles), GloSea5 ensemble means (squares) and 100 time-series of the variable $\mu_y + s_t$, drawn randomly from the Markov-Chain simulation (semi-transparent lines).}\label{fig:post-s}
\end{figure}

Having specified the prior distributions and the likelihood function, we now have all the ingredients to approximate the posterior distribution $p(\theta, s | x,y)$ by MCMC.
\reffig\ \ref{fig:mu-chain} shows 200 MCMC samples, and the estimated posterior distributions of the parameters $\mu_x$ and $\mu_y$.
 The posterior distributions of $\mu_x$ and $\mu_y$ were estimated from all $10^5$ MCMC samples.
The posterior distribution of $\mu_x$ is narrower than that of $\mu_y$ because the availability of 24 ensemble members allows for a more robust estimation of $\mu_x$ than $\mu_y$, which is only based on one observational time series.
 Both posterior distributions, of $\mu_x$ and $\mu_y$, have slightly heavier tails than the corresponding Normal distributions (not shown).
 The posterior means (standard deviations) are $23.4hPa$ ($0.56hPa$) for $\mu_x$ and $20.9hPa$ ($1.80hPa$) for $\mu_y$.
The model bias, defined by $\mu_x - \mu_y$ has posterior mean of $2.55hPa$ and posterior standard deviation of $1.64hPa$, resulting in a posterior probability of a positive bias $Pr(\mu_x > \mu_y) = 0.94$, and a posterior probability of $0.83$ that the bias exceeds $1hPa$.

\reffig\ \ref{fig:post-s} shows that MCMC approximation allows for estimation of the latent variable $s_t$, of which 100 samples from the Markov Chain are shown (shifted upward by $\mu_y$).
 The estimated time series of $s_t$ are used in \refsec\ \ref{sec:unc-corr} where we generate new artificial ensemble forecasts for the 1992--2011 NAO observations to quantify uncertainty in correlation coefficients.

\begin{figure}
\centering\includegraphics{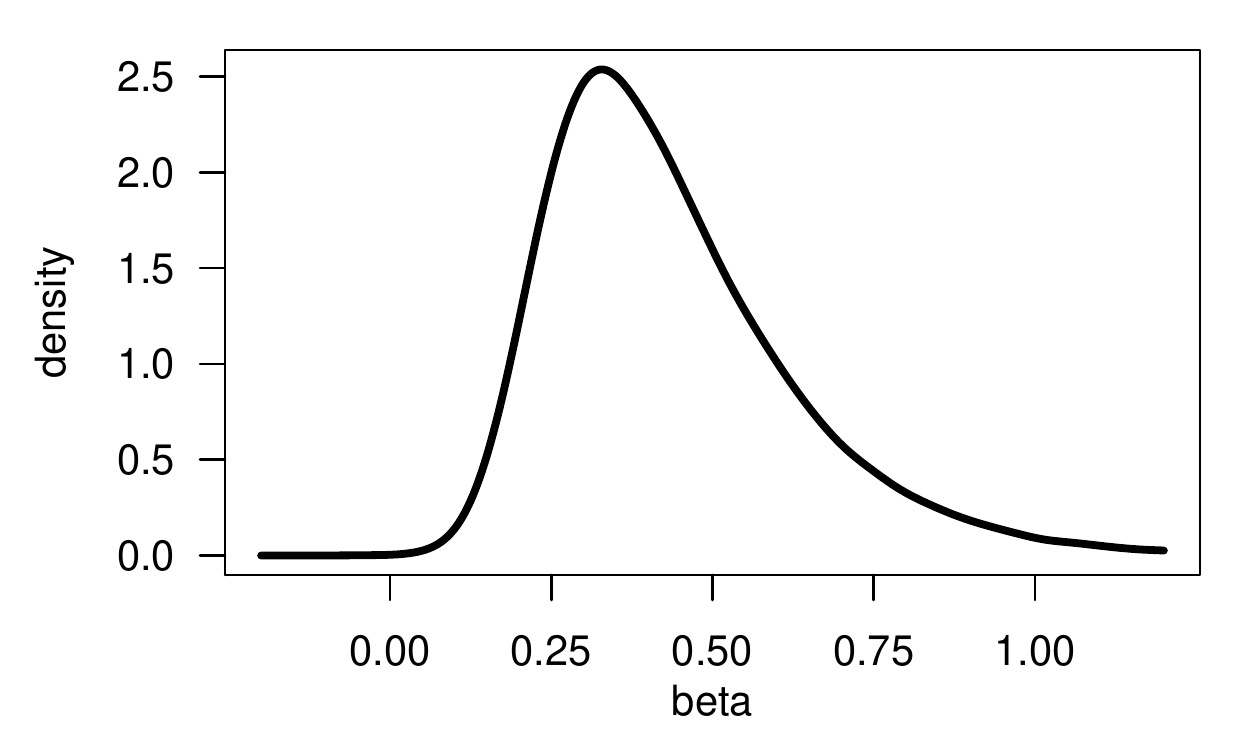}
\caption{Posterior distribution of $\beta$.}\label{fig:post-beta}
\end{figure}

The posterior distribution for $\beta$ is shown in \reffig\ \ref{fig:post-beta}.
The parameter $\beta$ quantifies how sensitive the forecasts are to the predictable signal relative to the sensitivity of the observations. 
When $\beta\neq 0$ there is dependency between forecast and observations - the forecasting system has skill. 
From the posterior distribution, the probability $Pr(\beta > 0) = 0.99$ 
and $Pr(\beta>0.2) = 0.95$, and so we are confident that the forecasting system has skill for predicting the NAO. 
But are the forecasts reliable, i.e., are the raw ensemble members exchangeable with the observations? 
A necessary condition for reliability of the raw forecasts is that $\beta=1$, which appears highly unlikely from our
posterior distribution, which gives $Pr(\beta<1)=0.99$ and $Pr(\beta<0.8)=0.95$. 
This means that individual raw forecasts should not be taken on face value as possible 
realizations of the observations, which is in agreement with the conclusions of \citet{eade2014seasonal}, and highlights that statistical recalibration of the raw forecasts is necessary.
Note that $\beta < 1$ implies that the model only contains a damped version of the predictable signal $s_t$. 
The posterior distribution of $\beta$ thus indicates an anomalously low signal-to-noise ratio of the ensemble, which we analyze in more depth in \refsec\ \ref{sec:sn-ana}.

\begin{figure}
\centering\includegraphics{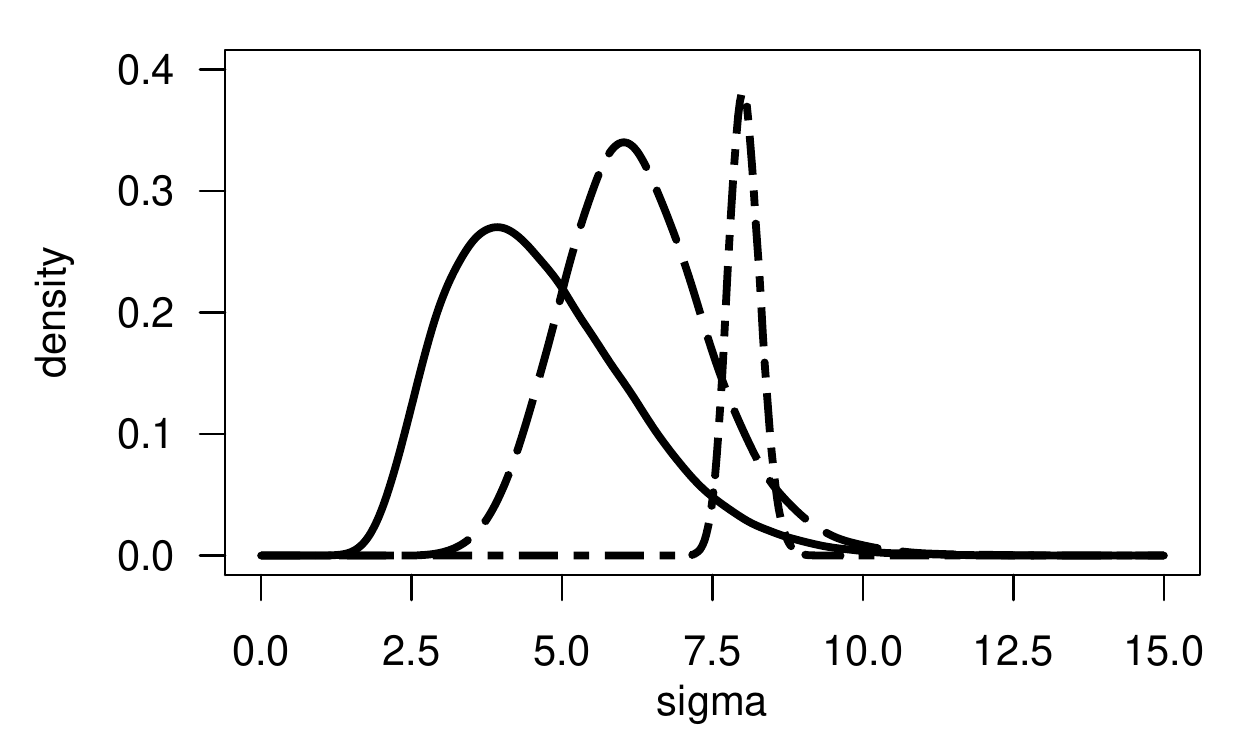}
\caption{Marginal posterior distributions of $\sigma_s$ (full line), $\sigma_\epsilon$ (dashed line), and $\sigma_\eta$ (dot-dashed line; scaled by $1/4$).}\label{fig:post-sigma}
\end{figure}

\reffig\ \ref{fig:post-sigma}  shows the posterior distributions of the parameters $\sigma_s$, $\sigma_\epsilon$, and $\sigma_\eta$.
 The posterior means (standard deviations) are $4.66hPa$ ($1.53hPa$) for $\sigma_s$, $6.26hPa$ ($1.22hPa$) for $\sigma_\epsilon$, and $8.03hPa$ ($0.26hPa$) for $\sigma_\eta$.
 It should be noted that $\sigma_s$ and $\sigma_\epsilon$ are highly dependent:
 According to \refeq\ (\ref{eq:vary}), the sum of their squares is constrained by the variance of the observations; the total variance of the observations can be explained either by lots of signal and little noise, or little signal and lots of noise.
 If only the observations were available, $\sigma_s$ and $\sigma_\epsilon$ would be unidentifiable.
 Only by basing the inference on the forecast system can $\sigma_s$ (and therefore $\sigma_\epsilon$) be constrained, however considerable uncertainty remains.
 In contrast to $\sigma_\epsilon$, the parameter $\sigma_\eta$ is better constrained by the data, because the individual ensemble members allow for estimation of the residual variance around the ensemble mean.
 A posterior comparison between the noise amplitudes yields $P(\sigma_\eta >
\sigma_\epsilon) = 0.92$, i.e., there appears to be more unpredictable noise in the
forecasting system than in the observations.
At the same time, there is good agreement between the total standard deviations of the observations and the individual ensemble members, as defined in \refeq\ (\ref{eq:varcov}): The posterior mean (standard deviation) is $7.97hPa$ ($1.09hPa$) for $\sqrt{var(y)}$, and $8.25hPa$ ($0.28hPa$) for $\sqrt{var(x_i)}$.

Note that the model parameters are not invariant under linear transformations 
of either the observations or forecasts. 
However, since the NAO is often defined in different ways (e.g., by the leading sea level 
pressure empirical orthogonal function, or station pressure difference, or area averaged pressure difference, and possibly transformed to a normalized climate index), it is desirable that forecast 
performance should be based on quantities that are invariant to choice of linear scale. 
We will therefore now focus on scale-invariant functions of the parameters, namely the 
correlation coefficient in \refsec\ \ref{sec:unc-corr}, and signal-to-noise ratios in \refsec\ \ref{sec:sn-ana}.

\subsection{Uncertainty in the correlation coefficient}\label{sec:unc-corr}

A widely used evaluation criterion for ensemble mean forecasts is the Pearson correlation coefficient between the ensemble forecasts and observations given by
\begin{equation}
r_{\bar{x}y} = \frac{s_{\bar{x}y}}{\sqrt{v_{\bar{x}}v_y}}.
\end{equation}
For the hindcast data presented in \refsec\ \ref{sec:data}, the sample correlation is $r_{\bar{x}y}=0.62$.
Uncertainty in correlation coefficients is usually quantified by confidence intervals and p-values \citep[][sec.\ 8.2.3]{storch2001statistical}.
This section presents a posterior analysis of uncertainty in the correlation coefficient of NAO hindcasts of \refsec\ \ref{sec:data}. We address three precise questions.
It should be noted that the approach outlined below is applicable to other performance measures, such as the mean squared error (MSE), the continuous ranked probability score (CRPS),  or the Ignorance score.

\paragraph{What is the uncertainty in the population correlation coefficient $\rho$, given the hindcast data?}

\begin{figure}
\centering\includegraphics{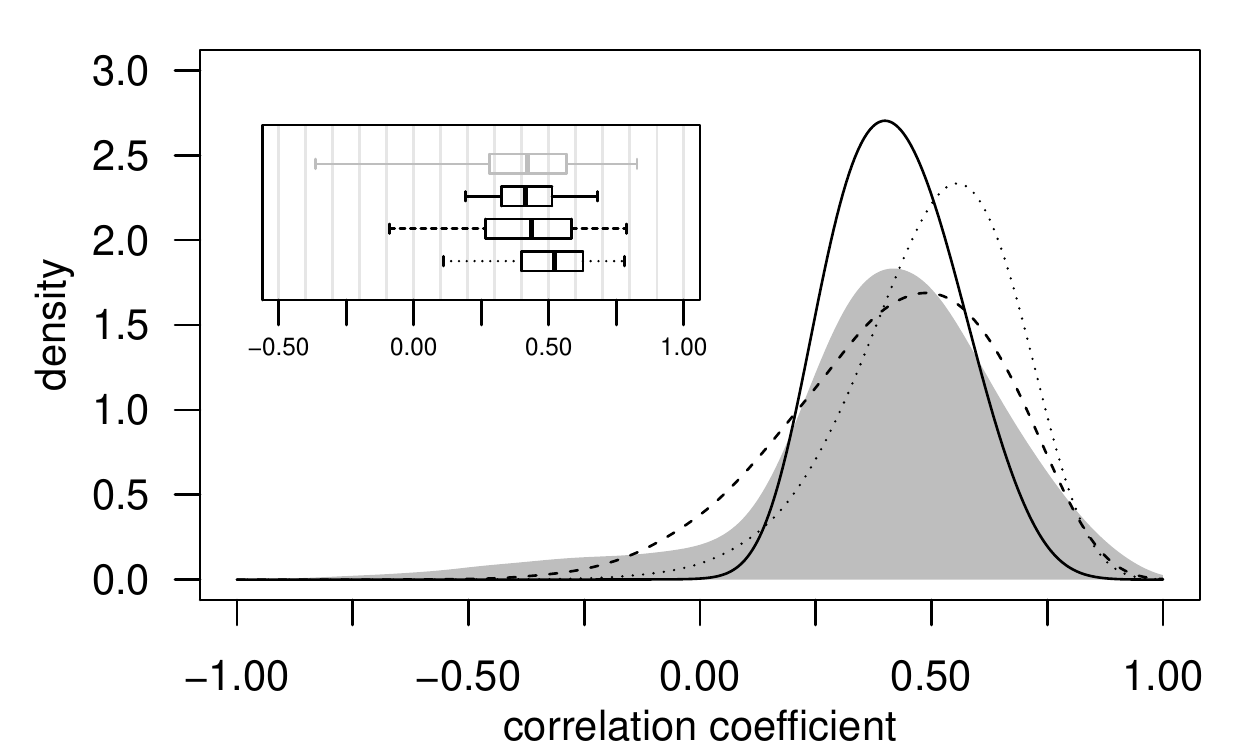}
\caption{Uncertainty in the correlation coefficient. Prior of the population correlation $\rho$(gray area), its posterior distribution (solid line), the posterior predictive distribution of the sample correlation over arbitrary 20-year periods (dashed line), and the posterior predictive distribution of the sample correlation with observations fixed at the actual 1992--2011 NAO values (dotted line). The Box-and-Whiskers plots in the inset indicate the 2.5, 25, 50, 75, and 97.5 percentiles of the distributions.}
\label{fig:corr-unc}
\end{figure}

In other words: What are possible values of the correlation coefficient taken over infinitely many 24-member ensembles and corresponding NAO observations, from which the given hindcast data is only a random sample of size $N=20$?
To answer this question, we consider the population correlation coefficient $\rho$ of the 24-member ensemble mean, expressed as a function of the model parameters, as given by \refeq\ (\ref{eq:rho-pop}).
We calculate $\rho$ for each MCMC sample of the model parameters, and thereby approximate the posterior distribution of the population correlation coefficient.
Our prior and updated posterior distribution of $\rho$ are indicated by the gray area and the solid line in \reffig\ \ref{fig:corr-unc}, respectively.
The posterior distribution of $\rho$ quantifies our uncertainty about the correlation coefficient due to uncertainty in the parameters of the statistical model, and due to the fact that the ensemble mean can only be estimated imperfectly by 24 ensemble members (thus the term $\sigma_\eta^2/R$ in the denominator of \refeq\ \ref{eq:rho-pop}).
 Due to the mode of the prior distribution of $0.4$, the posterior mean of $\rho$ of $0.42$ is smaller than the actual sample correlation of $0.62$.
 20 samples of hindcast data are not sufficient to override the prior too much, and the result is therefore biased towards our prior judgments about NAO skill. 
It might be argued that this result is unduly influenced by the prior distribution on the correlation.
In appendix D, we illustrate the sensitivity of the posterior distribution of the correlation for different prior distributions.
The sensitivity analysis shows that even for a very optimistic prior distribution, with prior mode at a correlation of $0.7$, the posterior mode of the correlation is shrunk down to about $0.5$.
The central $95\%$ credible interval derived from the posterior distribution (depicted in the inset in \reffig\ \ref{fig:corr-unc} is equal to $[0.19, 0.68]$, which does not overlap zero, but which also has the sample correlation value of $0.62$ in its upper tail.
In conclusion, the sample correlation coefficient of $0.62$ might be an overestimation of the true correlation skill of GloSea5, but we can say with high certainty that the system does have positive correlation skill.

\paragraph{What is the uncertainty in the sample correlation coefficient $r_{\bar{x}y}$ for different non-overlapping 20 year forecast periods?}
 To answer this question, we calculate a large collection of sample correlation coefficients $r_{\bar{x}y}$ as follows:
 We draw a set of parameters $\{\mu_x, \mu_y, \beta, \sigma_s,\sigma_\epsilon,\sigma_\eta\}$ from the MCMC output. 
 We use $\sigma_s$ to sample a random signal time series $s_1,\cdots,s_{20}$, and then use the other parameters to generate a random hindcast data set with $R=24$ and $N=20$ according to \refeq\ (\ref{eq:model}). 
 We then calculate the sample correlation $r_{\bar{x}y}$ of the ensemble mean in this artificial data set, and repeat this process for all $10^5$ MCMC samples.
 The resulting distribution is the posterior predictive distribution of $r_{\bar{x}y}$ (``predictive'' because it is a distribution over observables rather than parameters \citep{gelman2004bayesian}).
 The posterior predictive distribution is indicated by the dashed line in \reffig\ \ref{fig:corr-unc}.
 This distribution accounts for parameter uncertainty (because we sample parameters from the posterior), and also for finite-sample uncertainty (because we draw a random hindcast data set of finite length $N$).
 The posterior predictive distribution therefore quantifies our uncertainty about the sample correlation calculated over an arbitrary 20-year period.
 The posterior mean and median of this predictive distribution are very close to that of the posterior distribution of $\rho$.
 But the predictive distribution is wider than the posterior distribution.
The $95\%$ credible interval derived from this distribution is $[-0.09, 0.79]$.
 Taking into account finite sample uncertainty in addition to parameter uncertainty increases the overall uncertainty.

\paragraph{What is the uncertainty in the sample correlation coefficient $r_{\bar{x}y}$ for the same 1992--2011 NAO observations, but for a new realization of the ensemble forecast?}
 To answer this question, we calculate the posterior predictive distribution of $r_{\bar{x}y}$, where the observations are fixed at their values shown in \reffig\ \ref{fig:nao}. 
 That is, we generate replicated ensemble forecasts for these particular observations.
 To do this, we sample a signal time series $s_1,\cdots,s_{20}$ directly from the MCMC output (sketched in \reffig\ \ref{fig:post-s}), instead of generating $s_1,\cdots,s_{20}$ randomly.  
 We also draw the parameters $\beta$ and $\sigma_\eta$ from the same iteration of the Markov chain.
 We use these parameters to construct a new 24 member ensemble forecast using \refeq\ (\ref{eq:model-x}), and then calculate the sample correlation with the original 1992-2011 NAO observations.
 Note that the sampled series of $s_t$ is correlated with the original observations, and therefore the resampled ensemble members will be correlated with the original observations as well.
 The corresponding posterior predictive distribution of $r_{\bar{x}y}$ is indicated by the dotted line in \reffig\ \ref{fig:corr-unc}.
 Treating the observations as fixed quantities and only the ensembles as random decreases the width of the distribution, the $95\%$ credible interval is now $[0.11, 0.78]$.
Furthermore, the predictive mean and mode of this distribution are about $0.5$, i.e., slightly higher than the means and modes of the previous distributions.
Our best explanation for this shift is that it is caused by the last 3 NAO observations which represent large excursions from the mean compared to the previous 17 observations, and thereby bias the correlation coefficient upward compared to randomly sampled observations from a Normal distribution.
On the one hand this would imply that the Normal assumption is inadequate for the data. 
On the other hand, comparison of the two predictive distributions of $r_{\bar{x}y}$ (for fixed and arbitrary observational periods) suggests that in the future, when NAO might exhibit more normal behavior, the sample correlation using the same model will probably become smaller than 0.62.

\subsection{Signal-to-noise analysis}\label{sec:sn-ana}

It has been noted by \citet{scaife2014skillful}, \citet{kumar2014relationship} and \citet{eade2014seasonal} that the signal-to-noise ratios in seasonal climate predictions can be too low, which leads to the counter-intuitive effect that the ensemble forecasting system is less skillful at predicting members drawn from itself than at predicting the observation.
This is problematic because the skill of the ensemble at predicting itself is often assumed to be an upper bound of predictability of the real world.
Previous studies have provided only point estimates of signal-to-noise ratios and have not quantified how much uncertainty is in these quantities, which was criticized by \citet{shi2015impact}.
A Bayesian framework allows us to calculate posterior probabilities for hypotheses related to signal-to-noise ratios.

 In \citet{eade2014seasonal}, the ratio of predictable components ($\rpc$) was proposed as a measure to compare levels of predictability in the forecasting system and in the real world.
 The predictable component of the real world $\text{PC}_{\text{obs}}$ was defined as the correlation between the ensemble mean and the observations, and the predictable component of the model $\text{PC}_{\text{mod}}$ was defined as the ratio of the standard deviation of the ensemble mean and the mean standard deviation of the ensemble members. 
 $\rpc$ equals the ratio $\pcobs/\pcmod$, and was found by \citet{eade2014seasonal} to be about $2$ for the NAO hindcast.

 $\pcobs$, $\pcmod$, and $\rpc$, expressed in terms of the parameters of the signal-plus-noise model are given in appendix B.
 \citet{eade2014seasonal} argue that for a forecasting system that ``perfectly reflects the actual predictability'', $\rpc$ should be equal to one.
 If we define a perfect forecasting system by full exchangeability of ensemble members and observations, i.e., $\mu_x=\mu_y$, $\sigma_\epsilon=\sigma_\eta$, and $\beta=1$, and substituting these equalities into \refeq\ (\ref{eq:rpc}), we find that the perfect value of $\rpc$ is
\begin{equation}
\rpcperf = \left(1+\frac{1}{R\sigma_s^2/\sigma_\epsilon^2}\right)^{-1}.
\end{equation}
It can be noted from this that $\rpcperf\neq1$ even for a fully exchangeable system. 
To obtain $\rpcperf=1$, one also has to have either an infinitely large ensemble, i.e., $R\rightarrow\infty$, or no unpredictable noise in the system, i.e., $\sigma_\eta=\sigma_\epsilon=0$.
When both $R$ and $\sigma_\epsilon$ are finite, $\rpc$ is smaller than one.

\begin{figure}
\centering\includegraphics{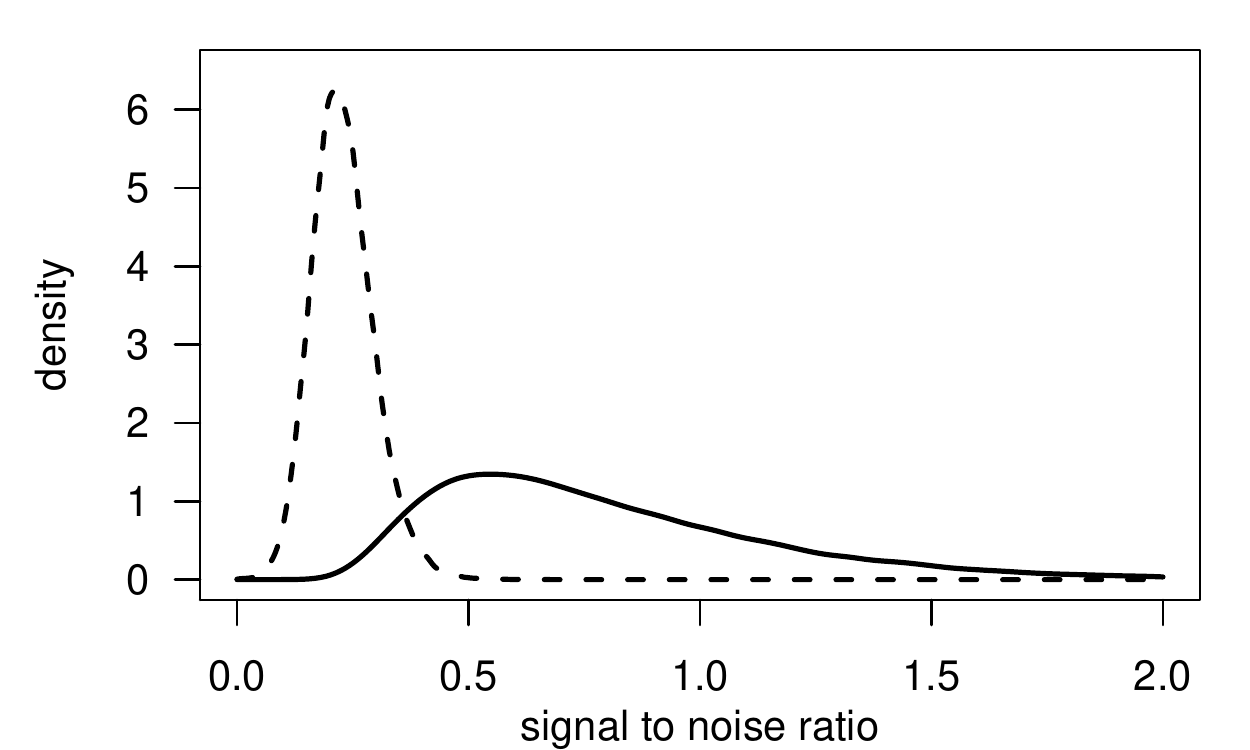}
\caption{Posterior distributions of the signal-to-noise ratio of the observations (full line) and of the model (dashed line).}\label{fig:post-snr}
\end{figure}

$\rpc$ is a rather complicated function of the $(\beta,\sigma_s,\sigma_\epsilon,\sigma_\eta)$ parameters (cf. \refeq\ \ref{eq:rpc}), and $\rpc=1$ corresponds to an imperfect forecasting system.
 Therefore, we shall consider instead  the signal-to-noise ratios ($\snr$) of the forecast system and of the observations.
The $\snr$'s are simply the ratio of the standard deviation of the predictable component (signal), and the unpredictable component (noise) of observations and of individual ensemble members, i.e.,
\begin{subequations}
\begin{align}
\snrobs & =\frac{\sigma_s}{\sigma_\epsilon}\text{, and}\\ 
\snrmod & = \frac{|\beta|\sigma_s}{\sigma_\eta}.
\end{align}
\label{eq:snr}
\end{subequations}
 Note that $\snrobs$ and $\snrmod$ are invariant under a shift or rescaling of the forecasts or the observations.
 Substituting the moment estimators from appendix C into \refeq\ (\ref{eq:snr}), we obtain $\snrobs=1.73$ and $\snrmod=0.21$, i.e., the observations appear to be more predictable than the model.
 But the model parameters are very uncertain.
 Therefore we should also expect $\snr$'s to be very uncertain.

\reffig\ \ref{fig:post-snr} shows posterior distributions of $\snrobs$ and $\snrmod$
derived from the MCMC simulation.  The posterior distribution of $\snrmod$ is
sharper than that of $\snrobs$ because 24 ensemble members allow for more
robust estimation than a single observation time series.  We confirm with very
high posterior probability the previous result of \citet{scaife2014skillful}
that, for the GloSea5 winter NAO forecast, the $\snr$ of the model is lower
than the $\snr$ of the observations.  In particular, we have a posterior
probability $Pr(\snrobs > \snrmod) = 0.99$ (updated from prior probability of
$\approx 0.5$). The sensitivity of this conclusion to the choice of the prior is briefly discussed in appendix D.

Our posterior analysis assigns very high probability to the hypothesis that the
predictable signal component in the model is weaker than in the real world. The
analysis of \citet{shi2015impact}, which is based on a set of winter NAO
hindcasts produced by different models, concludes that such an underconfident
ensemble ``merely suggest an inadequately small sample size''.  Contrary to
that, the analysis based on our 20-year data set (and our statistical
assumptions) confirms the finding of \citet{eade2014seasonal} with very high
confidence, despite the small sample size: The raw GloSea5 ensemble
underestimates the predictability of the real world, and statistical
post-processing of the raw ensemble is necessary to generate reliable
forecasts.

\subsection{Calibration and prediction}\label{sec:cali-pred}

Bayesian inference using the signal-plus-noise model provides a natural framework for recalibrating forecasts
to produce reliable probability distributions of future observations. 
The predictive distribution function for the unknown observation $y_t$ is 
the conditional distribution of $y_t$, given the known quantities $\{x,y\}_{-t}$, i.e., the hindcast data set not including the time instance $t$, as well as $x_t$, i.e., the ensemble forecast for $y_t$.
The predictive distribution can be calculated by integrating over the posterior distribution of the model parameters:
\begin{equation}
p(y_t | \{x,y\}_{-t}, x_t) = \int d\theta\ p(y_t|x_t, \theta) p(\theta|\{x,y\}_{-t}, x_t).\label{eq:predictive}
\end{equation}
Note that according to \refeq\ (\ref{eq:linreg-post}), the conditional distribution $p(y_t|x_t,\theta)$ is a Normal distribution whose parameters depend on the signal-plus-noise model parameters.
The predictive distribution \refeq\ (\ref{eq:predictive}) can thus be interpreted as a weighted mixture of Normal distributions, where the weight is given by the posterior density of the model parameters.
A mixture of Normal distributions is itself not in general a Normal distribution. 
We should thus not expect the predictive distributions to be Normal, even though our statistical model is based on the assumption of Normality of the data.
 The resulting predictive distributions include a suitable predictive variance that takes into account parameter uncertainties and forecast uncertainty.
 We generate $N=20$ predictive distributions in leave-one-out mode, that is for each $t=1,\cdots,N$, the predictive distribution for $y_t$ is calculated under the assumption that $y_t$ is unknown (see \citet{friedman2001elements} for further details on cross validation).
The STAN code has to be adjusted slightly to simulate these out-of-sample predictive distributions (see appendix A).

 We compare the posterior predictive distribution functions to a simple benchmark given by ordinary linear regression.
 Recall that we have argued in \refsec\ \ref{sec:linreg-corr} that if the model parameters were known, linear regression would be the optimal post-processing method for signal-plus-noise models.
 We regress the observations $y_t$ on the ensemble means $\bar{x}_t$, and predict a Gaussian forecast distribution with the residual variance of the regression, i.e.,
\begin{equation}
(y_t | \bar{x}_t) \sim \mathcal{N}\left(m_y + \frac{s_{\bar{x}y}}{v_{\bar{x}}}(\bar{x}_t - m_x), v_y\left(1-r_{\bar{x}y}^2\right)\right)
\end{equation}
The benchmark predictions were generated in leave-one-out mode as well.

\begin{figure}
\centering\includegraphics{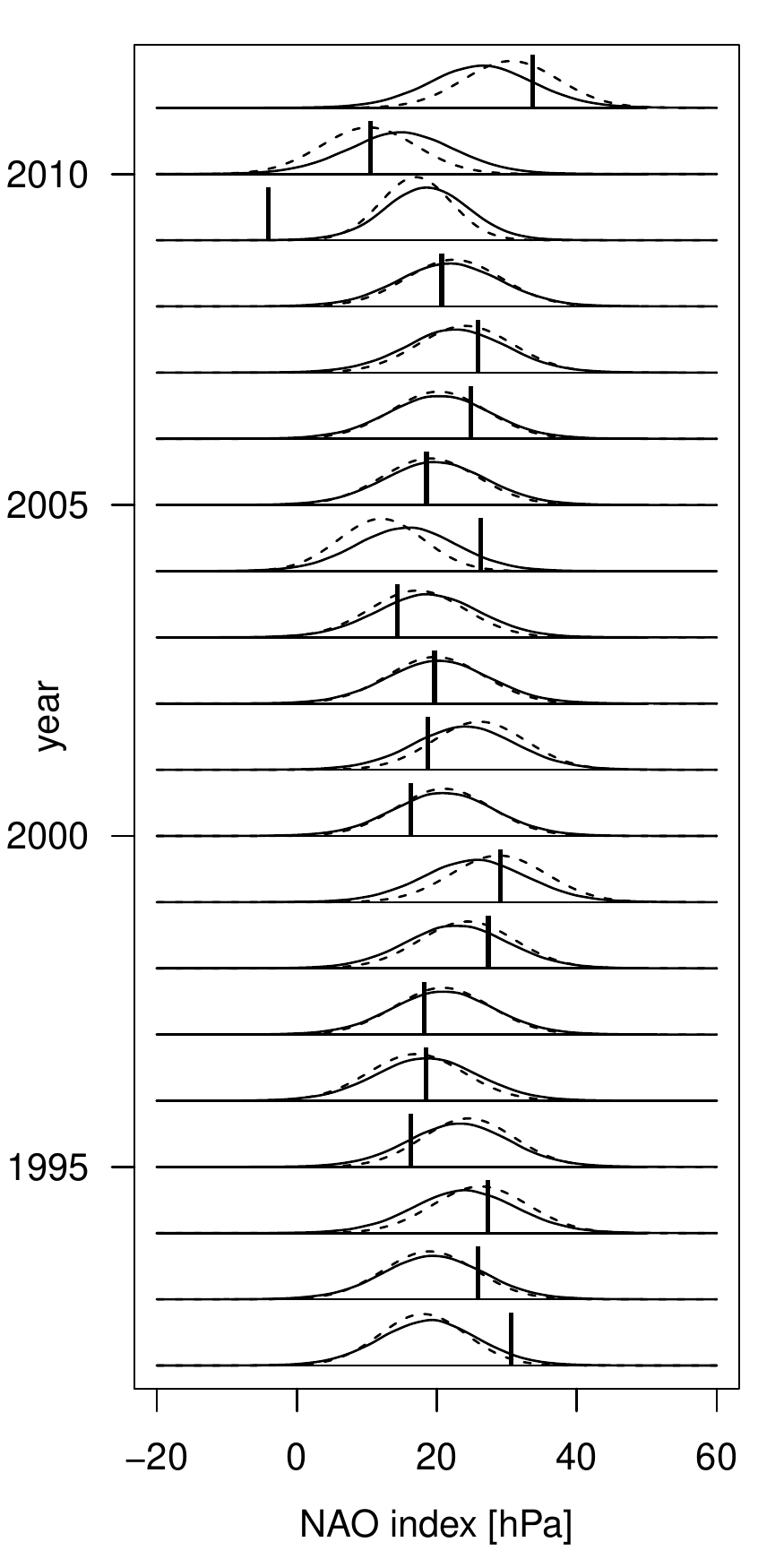}
\caption{Predictive distributions functions based on simple  linear regression (dashed line) and fully Bayesian, using the signal-plus-noise model (full lines). The vertical lines indicate the observations.}\label{fig:pred-pdfs}
\end{figure}

 The posterior predictive distributions and the benchmark predictions are shown in \reffig\ \ref{fig:pred-pdfs}.
 In general the posterior predictive distributions are wider than the benchmark predictions; their average standard deviations are $7.5hPa$ and $6.6hPa$, respectively.
 The posterior predictive means are less variable than the benchmark means; their standard deviations are $3.0hPa$ and $5.1hPa$, respectively.

 The larger dispersion of the posterior predictive distributions leads to the effect that in the majority of cases (when the NAO is close to its climatological mean) the benchmark predictions assign a higher predictive density to the observation than the posterior predictive distributions.
 On the other hand, if the observation is far away from the climatological mean, or far away from the forecast mean, the posterior predictive distributions assign more density to the observations.
 We address the question of which collection of forecasts are ``better'' on average by calculating the average Ignorance score \citep{roulston2002evaluating}.
 Given a forecast density $p(z)$  and a verifying observation $y$, the Ignorance score is defined by
\begin{equation}
\mathcal{I}(p; y) = -\log_2 p(y).
\end{equation}
 The Ignorance is a proper scoring rule for probabilistic forecasts of
continuous quantities; its average can be taken as a summary of forecast
performance, indicating better forecasts by lower values.

\begin{table}
\caption{Average ignorance scores and standard errors for different forecast methods}\label{tab:ign}
\begin{center}
\begin{tabular}{l c c}
Method & mean Ign. & standard error \\
\hline
climatology &  $5.46$ & $0.62$ \\
regression benchmark & $5.24$ & $0.62$ \\
posterior predictive & $5.02$ & $0.41$ 
\end{tabular}
\end{center}
\end{table}

 In \reftab\ \ref{tab:ign} we compare the average Ignorance scores of three different forecasts:
 The leave-one-out climatological forecast which is simply a Normal distribution with the climatological mean and variance, the linear regression benchmark, and the posterior predictive distributions.
 It is reassuring that the posterior predictive distributions assign a higher average density to the observation than both, the climatology and the regression benchmark.
 The additional skill is due to the wider predictive distributions and the less variable predictive mean.
 These two features are a consequence of accounting for parameter uncertainty by integrating over their posterior distribution.
 In conclusion, the Bayesian analysis using a signal-plus-noise model not only provides useful evaluation diagnostics, but also provides a natural way of generating skillful and well-calibrated probability forecasts.

\section{Discussion}

\subsection{Model criticism}

We have used a simplified statistical model to make inference about an actual forecasting system, so it is important to be aware of the limitations of the statistical model.
It is important not to confuse limitations of our statistical model with deficiencies of the real forecasting system.

There are a number of features of observed climate indices and their ensemble forecasts that our simplified model cannot account for. These include:
Autocorrelation in the ensemble forecasting system and the observations;
a spread-skill relation, that is, a systematic relationship between the ensemble spread and the distance between the ensemble mean and the verifying observation;
trend in the observations and drifts in the model output;
skewness, bimodality, or heavy-tailedness of the distribution of the predictand.
 More work is necessary to develop statistical frameworks for ensemble forecasts that take some or all of these effects into account without becoming overly complex. 
 On the other hand, by leaving out all these details, our model retains a high level of interpretability.
 Before making the model more complex we also have to ask ourselves, how much information can we justifiably hope to infer from 20 years' worth of annual hindcast data?

\subsection{Model checking}\label{sec:model-checking}

We have tested the validity of our exchangeability assumptions in \refsec\ \ref{sec:sn-model-sub} by replacing the observation by one of the ensemble members.
Since we judged the ensemble members to be exchangeable, replacing the observation by an ensemble member should produce a perfect model scenario, where the observation and ensemble members are statistically indistinguishable from each other, i.e., we should have $\mu_x=\mu_y$, $\beta=1$, $\sigma_\epsilon=\sigma_\eta$.
After rerunning the posterior analysis under this perfect model scenario, we found that the posterior distributions of $\mu_x$ and $\mu_y$, and of $\sigma_\epsilon$ and $\sigma_\eta$ overlap each other and provide no indication for non-exchangeability. 
Furthermore, the posterior distribution of $\beta$ does not rule out the value $\beta=1$ as strongly as the posterior distribution shown in \reffig\ \ref{fig:post-beta}.
However, we still found the bulk of the posterior distribution of $\beta$ to be concentrated between 0 and 1, resulting in a rather high posterior probability of $Pr(\beta<1)\approx0.95$.
Furthermore, we found a posterior probability for an anomalous signal-to-noise ratio of $Pr(\snrobsperf > \snrmod) \approx 0.85$ in this perfect-model scenario.
These posterior probabilities provide evidence that our statistical model might not be flexible enough to accurately model the data.
A possible explanation for the observed behavior is that the ensemble members are, in fact, not
exchangeable with each other, which could be the result of the lagged initialization of GloSea5 on 3 different dates.
Without further analyses we are unable to model such an ensemble with non-exchangeable members and we leave this problem open for future studies.

We have also repeated our analyses with different NAO observations, taken directly from station data at Lisbon and Reykjavik \citep{hurrell2014station}, and from the leading empirical orthogonal function of sea level pressure anomalies over the Atlantic sector \citep{ncar2014pc}.
The posterior distributions of the $\mu$'s and $\sigma$'s change slightly because the alternative observations have different scales.
For the scale-invariant quantities analyzed in \refsec\ \ref{sec:unc-corr} and \ref{sec:sn-ana}, however, the posterior distributions are 
almost identical to the ones we obtained earlier. Our main conclusions are therefore insensitive to the choice of NAO observations.

\subsection{Correlation uncertainty under different hindcast settings} 

\begin{figure}
\centering\includegraphics{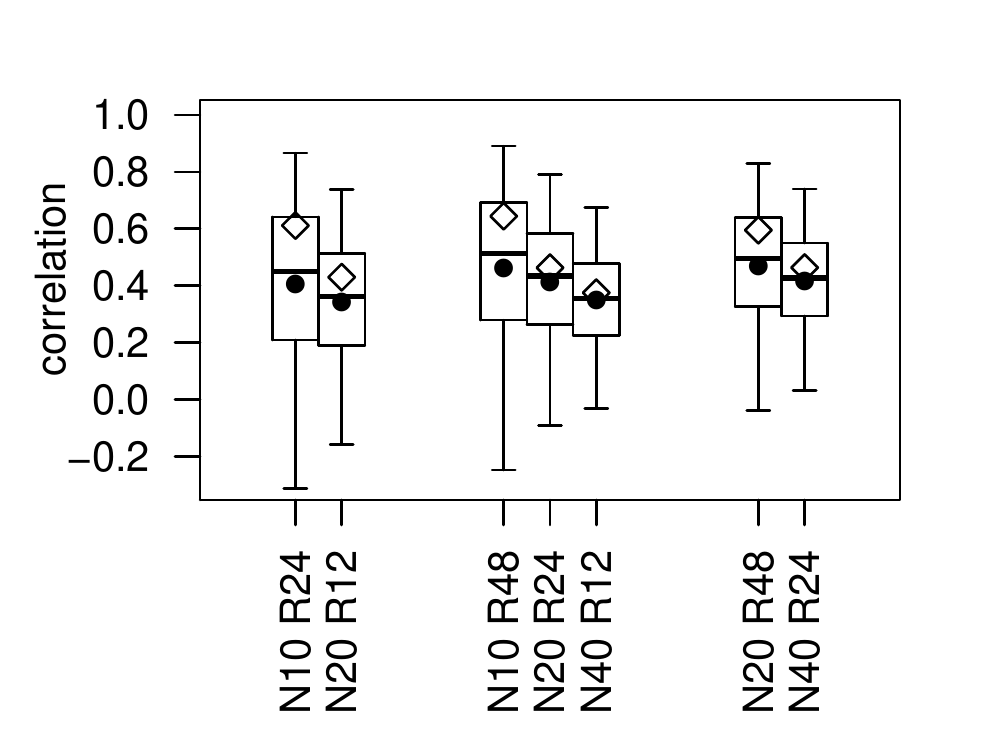}
\caption{Posterior predictive distribution of correlation coefficients for different combinations of $N$ and $R$. The box-and-whiskers indicate the 2.5, 25, 50, 75, and 97.5 percentiles of the predictive distributions, the diamonds indicate the mode and the circles the mean. The plots are grouped according to their computational expense $N\times R$.}\label{fig:corr-comp-exp}
\end{figure}

Statistical inference using a signal-plus-noise model might be useful for design of future ensemble systems. 
Simulations from the model can be used to calculate predictive distributions of correlation coefficients for different ensemble sizes $R$ and for different sample sizes $N$.
 In practice, the hindcast length $N$ and the ensemble size $R$ can usually not be chosen independently, but their choice is constrained by the available computational resources.
 Given that the computational expense of a planned hindcast experiment, defined by the product $NR$, is fixed, how should $N$ and $R$ be chosen?
 One possible criterion might be to consider the range of possible values of the correlation coefficient.
 \reffig\ \ref{fig:corr-comp-exp} shows that for a given computational expense (i.e., $NR$ constant), there is a trade off between mean and spread of the distribution
of possible correlation values. 
 Higher expected correlation can be obtained by increasing the ensemble size $R$
while decreasing the hindcast length $N$. 
 At the same time, however, the risk of getting very low sample correlations (e.g., not significantly different from 0) increases if $N$ is decreased.
 This is because the spread of possible correlation values becomes wider, but also because the larger $N$ is, the smaller will be the correlation values that are deemed ``significant'' by statistical tests.

\section{Conclusion}

This study has shown how a statistical model can be used to diagnose and improve the skill and reliability of an ensemble forecasting system.
The distributions-oriented approach \citep{murphy1987general} provides a complete summary of the forecasting system and observations using a signal-plus-noise model, whose parameters can be estimated by Bayesian inference.
Posterior distributions of the parameters can be used to simulate properties of any desired performance measure and its uncertainty under hypothetical designs of the ensemble forecasting system. 
The framework provides a straightforward method for calculating calibrated probability forecasts for future observables for a given set of ensemble forecasts. 

We conclude by revisiting the 5 questions specified in the introduction which guided the analysis of NAO hindcasts produced by the GloSea5 seasonal climate prediction system.
Question 1: There is indeed much sampling uncertainty in the correlation between the ensemble mean and observations. But there is also strong evidence of actual positive correlation skill: the 95\% credible interval of $[0.19,0.68]$ does not overlap zero.
Question 2: Our analysis suggests that very different correlation skill might be observed over different 20 year periods. In particular, the value of 0.62 is in the upper tail of the correlation distribution, suggesting a high chance of a decrease in correlation skill if GloSea5 were evaluated over different periods.
Question 3: The skill uncertainty over the same 1992--2010 evaluation period is smaller than over arbitrary 20 year evaluation periods. Our results suggest that the 20 year period is unusual and produces higher than normal correlation skill. The reasons for this are not entirely clear, but might be related to large deviations of NAO in the years 2008--2010.
Question 4: Forecasts are certainly not exchangeable with the observations and can therefore benefit from recalibration. A particular feature of non-exchangeability is the anomalous signal-to-noise ratio (SNR). We show with over $99\%$ posterior probability that the SNR is smaller in the model than in the observations, i.e. the predictable signal in the model is too weak.
Question 5: The probabilistic framework used in this study allows us to derive a recalibrated predictive distribution, i.e. the conditional distribution of the observation, given the ensemble forecast. We found that the Bayesian method of integrating over the parameter uncertainty distribution improves the forecast skill compared to a simpler recalibration method.

It is worthwhile to highlight a few important advantages of a Bayesian framework over more traditional approaches.
Firstly, the proposed statistical model is based on explicit assumptions, which creates transparency in how we interpret the observed data, and about how we think forecasts are related to the real world. 
Transparency is the basis for critically discussing assumptions, and revising these assumptions if necessary.
Secondly, all the analyses to answer our research questions are coherently based on the exact same assumptions about the data.
There are established methods to address each of our research questions in isolation, for example a t-test for the correlation coefficient \citep{storch2001statistical}, analysis of ratio of predictable components \citep[RPC, ][]{eade2014seasonal} to address signal-to-noise ratio, and non-homogeneous Gaussian regression \citep[NGR, ][]{gneiting2005calibrated} for forecast recalibration.
But these methods are not explicitly based on the same statistical assumptions.
An explicit statistical model allows us to address different questions in a coherent way, without changing our assumptions about the data.
Lastly, uncertainty quantification is a crucial aspect of analyzing small climate hindcast data sets.
 In Bayesian analyses, probability is the primitive quantity, and uncertainty quantification is therefore built into the analysis by default.
 All questions can be addressed by posterior probability distributions, which not only communicate our best guesses, but also our degree of uncertainty.
On the other hand, computational methods for Bayesian analyses can be expensive, the specification of suitable prior distributions is problematic, and all conclusions are conditional on the parametric model assumptions being correct.

In future studies, it will be of interest to relax model assumptions (e.g., to include serial dependence in the signal time series), and to extend the model to allow for possible sources of non-stationarity (e.g., climate change trends), as well as spread-skill relationships. 
A more disciplined way of specifying the prior distribution over model parameters is needed.
It will also be of interest to develop computationally efficient methods for modeling spatial ensemble hindcasts and observations available at many grid point locations.

\section*{Acknowledgments}
We wish to thank Theo Economou, Ben Youngman and Danny Williamson for valuable discussions on Bayesian theory and MCMC. Further comments from Peter Watson and 3 anonymouns reviewers greatly improved the paper. This work was supported by the European Union Programme FP7/2007-13 under grant agreement 3038378 (SPECS), and by the joint DECC/Defra Met office Hadley Centre Climate Programme (GA1101).

\begin{appendix}

\section*{Appendix A: STAN model code}\label{app:stan}

For the diagnostic analysis, where all $N$ observations and ensemble forecasts are known, the following STAN code was used to approximate the posterior distribution:

\lstset{basicstyle=\ttfamily,keepspaces=true,columns=flexible}
\begin{lstlisting}
data {
  int<lower=1> N; 
  int<lower=1> R; 
  matrix[N,R] x; 
  vector[N] y; 
}
parameters {
  real mu_x;
  real mu_y;
  real<lower=0> sigma2_eps;
  real<lower=0> sigma2_eta;
  real<lower=0> sigma2_s;
  real beta;
  vector[N] s;
}
model {
  mu_x ~ normal(0, 30);
  mu_y ~ normal(0, 30);
  beta ~ normal(1, 0.7);
  sigma2_s ~ inv_gamma(2, 25);
  sigma2_eps ~ inv_gamma(3, 100);
  sigma2_eta ~ inv_gamma(3, 100);
  s ~ normal(0, sqrt(sigma2_s));

  y ~ normal(mu_y + s, sqrt(sigma2_eps));
  for (n in 1:N)
    for (r in 1:R)
      x[n,r] ~ normal(mu_x + beta * s[n], sqrt(sigma2_eta));
}
\end{lstlisting}

In order to generate the predictive distributions for \refsec\ \ref{sec:cali-pred}, where the $N$-th observation is assumed to be unknown, the following STAN code was used:

\begin{lstlisting}
data {
  int<lower=1> N;
  int<lower=1> R;
  matrix[N,R] x;
  vector[N-1] y;
}
parameters {
  real mu_x;
  real mu_y;
  real<lower=0> sigma2_eps;
  real<lower=0> sigma2_eta;
  real<lower=0> sigma2_s;
  real beta;
  vector[N-1] s;
  real s_new;
}
model {
  mu_x ~ normal(0, 30);
  mu_y ~ normal(0, 30);
  beta ~ normal(1, 0.7);
  sigma2_s ~ inv_gamma(2, 25);
  sigma2_eps ~ inv_gamma(3, 100);
  sigma2_eta ~ inv_gamma(3, 100);
  s ~ normal(0, sqrt(sigma2_s));

  y ~ normal(mu_y + s, sqrt(sigma2_eps));
  for (n in 1:(N-1)) 
    for (r in 1:R)
      x[n,r] ~ normal(mu_x + beta * s[n], sqrt(sigma2_eta));

  s_new ~ normal(0, sigma_s);
  for (r in 1:R)
      x[N,r] ~ normal(mu_x + beta * s_new, sqrt(sigma2_eta));
}
generated quantities {
  real y_new;
  y_new <- normal_rng(mu_y + s_new, sqrt(sigma2_eps));
}
\end{lstlisting}

\section*{Appendix B: Ratio of predictable components as functions of model parameters}\label{app:rpc}

This appendix complements \refsec\ \ref{sec:sn-ana}.
$\pcobs$, $\pcmod$, and $\rpc$, expressed in terms of the parameters of the signal-plus-noise model are given by
\begin{subequations}
\begin{align}
\pcobs & = \frac{\beta\sigma_s^2}{\sqrt{(\sigma_s^2+\sigma_\epsilon^2)(\beta^2\sigma_s^2+\sigma_\eta^2/R)}},\\
\pcmod & = \sqrt{\frac{\beta^2\sigma_s^2 + \sigma_\eta^2/R}{\beta^2\sigma_s^2+\sigma_\eta^2}}\text{, and}\\
\rpc & = \sqrt{\frac{1+\sigma_\eta^2/(\beta^2\sigma_s^2)}{\left(1+\sigma_\epsilon^2/\sigma_s^2\right)\left(1+\sigma_\eta^2/(R\beta^2\sigma_s^2)\right)}}.\label{eq:rpc}
\end{align}
\end{subequations}

\section*{Appendix C: Method of moment estimators for the signal-plus-noise model}\label{app:moment}

To calculate moment estimators for the parameters of the signal-plus-noise model \refeq\ (\ref{eq:model}), we use the summary measures given in \reftab\ \ref{tab:data-summary}, and additionally the average ensemble variance 
\begin{equation}
v_{x} = (NR)^{-1} \sum_{t=1}^N \sum_{r=1}^R (x_{t,r} - \bar{x}_t)^2.
\end{equation}

Equating the analytical first and second moments (cf. \refeq\ \ref{eq:varcov}) of the signal-plus-noise model with sample moments, and solving for the model parameters, we obtain estimating equations for the model parameters. The equations and corresponding values for the NAO data of \refsec\ \ref{sec:data} are summarised in \reftab\ \ref{tab:moment-est}.

\begin{table}[hbtp]
\caption{Estimating equations for parameters in the signal-plus-noise model derived by method of moments, and estimated values for the GloSea5 NAO hindcast data.}\label{tab:moment-est}
\begin{center}
\begin{tabular}{ccl|ll}
\multicolumn{3}{c|}{Estimating equations} & \multicolumn{2}{c}{Values}\\
\hline
$\hat{\mu}_x$ & $=$ & $m_x$ & $23.42$ & $hPa$\\
$\hat{\mu}_y$ & $=$ & $m_y$ & $20.94$ & $hPa$\\
$\hat{\sigma}_\eta^2$ & $=$ & $v_{x}$ & $62.17$ & $(hPa)^2$\\
$\hat\beta$ & $=$ & $s_{\bar{x}y}^{-1}(v_{\bar{x}} - R^{-1}v_{x})$  & $0.23$ &\\
$\hat\sigma_s^2$ & $=$ & $\hat\beta^{-1}s_{\bar{x}y}$ & $50.35$ & $(hPa)^2$\\
$\hat\sigma_\epsilon^2$ & $=$ & $v_{y} - \hat\sigma_s^2$& $16.77$ & $(hPa)^2$
\end{tabular}
\end{center}
\end{table}

\section*{Appendix D: Sensitivity to choice of priors}\label{app:sensitivity}

\begin{figure}
\centering\includegraphics{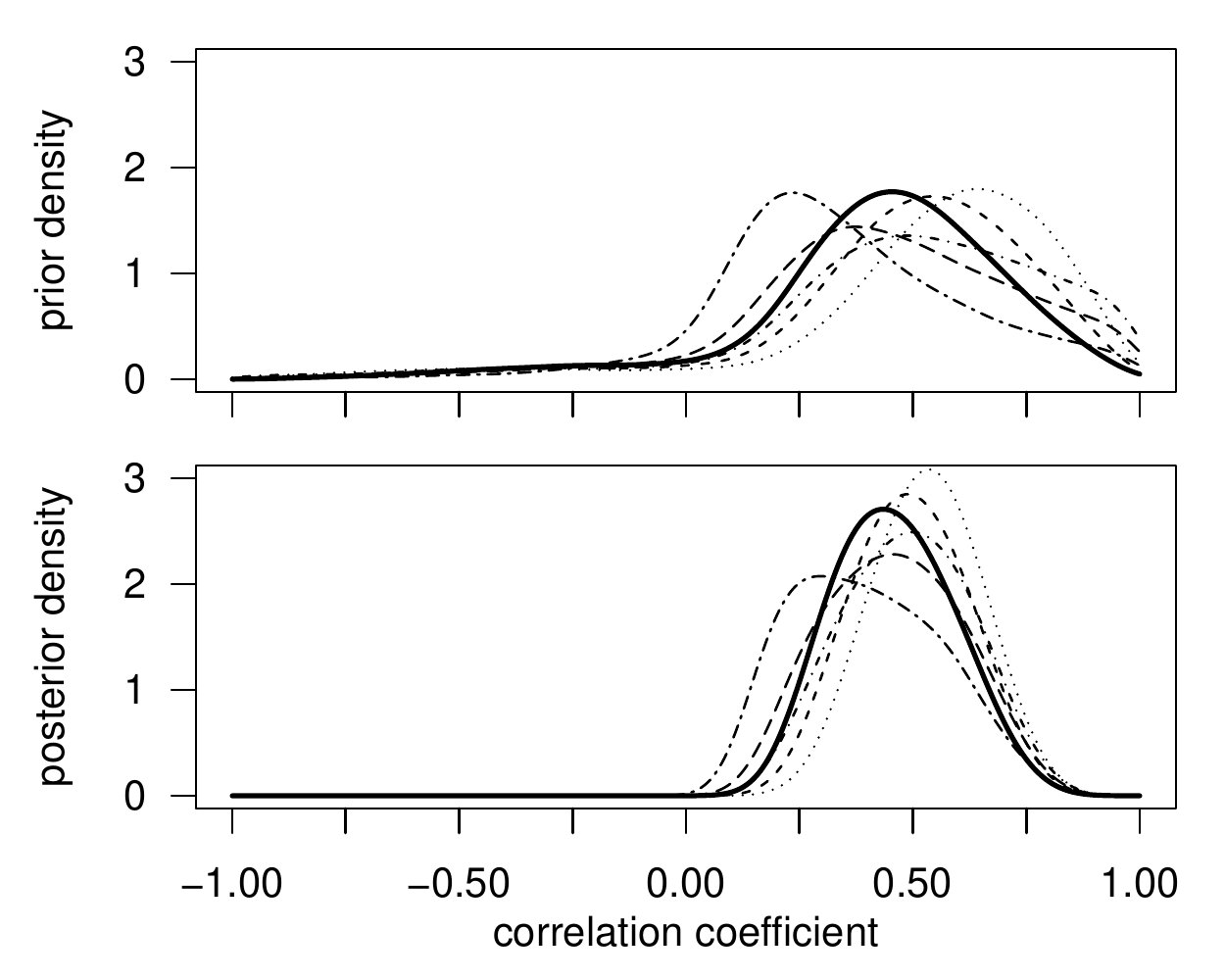}
\caption{Different prior distributions on the correlation coefficient (top panel) yield different posterior distributions (lower panel, with corresponding line types). The bold solid lines corresponds to the specifications of \refsec\ \ref{sec:bayes-updating}}.\label{fig:prior-sens}
\end{figure}

Bayesian analyses are sensitive to the choice of prior distributions.
This is desired for the present study; we want our prior judgments about diagnostic quantities to have an impact on our conclusions, especially because the sample size is small.
In order to illustrate the sensitivity to the choice of prior distributions, we consider the variability of the posterior distribution of the correlation coefficient $\rho$ (cf. \refsec\ \ref{sec:unc-corr} and \refeq\ \ref{eq:rho-pop}) when the prior parameters are varied.
We found the shape of the prior distribution on $\rho$ to be sensitive to the shape and scale parameters of the Inverse-Gamma prior distribution on $\sigma_s^2$.
We have varied these parameters within values that produce believable prior distributions on $\rho$. 
We have then calculated new posterior distributions of $\rho$ using the alternative prior specifications.
The varying prior distributions and updated posterior distributions of $\rho$ are shown in \reffig\ \ref{fig:prior-sens}.
As expected, due to the small sample size the posterior distributions vary considerably due to the variability of the prior.
But the differences between the different prior distributions is greater than the differences between their updated posterior distributions. 
Bayesian updating leads to a consensus between differing prior judgments.
Note further that the optimistic prior distributions with prior mode at $\approx
0.7$ is shrunk towards a mode at $\approx 0.5$, which is smaller than the
sample correlation of $0.62$ for the data.

In \refsec\ \ref{sec:sn-ana} we have shown that there is a
high posterior probability of an anomalously low signal-to-noise ratio of the
model: $Pr(\snrobs > \snrmod) > 0.99$. This probability is sensitive to the
choice of the prior parameters. Note that by changing only the prior
distribution of $\sigma_s^2$, the prior distribution of the correlation
changes, but the prior probability $Pr(\snrobs > \snrmod) \approx 0.5$ does not
change. We found that by changing the prior of $\sigma_s^2$ in such a way that
the correlation prior becomes more pessimistic, the posterior probability
for $(\snrobs > \snrmod)$ decreases. For the prior specifications that yield
the most pessimistic prior expectation of the correlation of $\approx 0.3$ in
\reffig\ \ref{fig:prior-sens}, the posterior probability of an anomalous SNR
reduces to $\approx 0.85$.

\end{appendix}

\bibliographystyle{abbrvnat}
\bibliography{bib}

\begin{thebibliography}{52}
\providecommand{\natexlab}[1]{#1}
\providecommand{\url}[1]{\texttt{#1}}
\expandafter\ifx\csname urlstyle\endcsname\relax
  \providecommand{\doi}[1]{doi: #1}\else
  \providecommand{\doi}{doi: \begingroup \urlstyle{rm}\Url}\fi

\bibitem[Annan and Hargreaves(2010)]{annan2010reliability}
J.~Annan and J.~Hargreaves.
\newblock {Reliability of the CMIP3 ensemble}.
\newblock \emph{Geophysical Research Letters}, 37\penalty0 (2), 2010.

\bibitem[Bartholomew et~al.(2011)Bartholomew, Knott, and
  Moustaki]{bartholomew2011latent}
D.~J. Bartholomew, M.~Knott, and I.~Moustaki.
\newblock \emph{{Latent Variable Models and Factor Analysis: A Unified
  Approach}}, volume 899.
\newblock John Wiley \& Sons, 2011.

\bibitem[Bradley et~al.(2004)Bradley, Schwartz, and
  Hashino]{bradley2004distributions}
A.~A. Bradley, S.~S. Schwartz, and T.~Hashino.
\newblock {Distributions-oriented verification of ensemble streamflow
  predictions}.
\newblock \emph{Journal of Hydrometeorology}, 5\penalty0 (3):\penalty0
  532--545, 2004.

\bibitem[Brooks et~al.(2011)Brooks, Gelman, Jones, and
  Meng]{brooks2011handbook}
S.~Brooks, A.~Gelman, G.~Jones, and X.-L. Meng.
\newblock \emph{{{Handbook of Markov Chain Monte Carlo}}}.
\newblock CRC Press, 2011.

\bibitem[Buonaccorsi(2010)]{buonaccorsi2010measurement}
J.~P. Buonaccorsi.
\newblock \emph{{Measurement Error: Models, Methods, and Applications}}.
\newblock CRC Press, 2010.

\bibitem[Chandler(2013)]{chandler2013exploiting}
R.~E. Chandler.
\newblock {Exploiting strength, discounting weakness: Combining information
  from multiple climate simulators}.
\newblock \emph{Philosophical Transactions of the Royal Society A:
  Mathematical, Physical and Engineering Sciences}, 371\penalty0 (1991), 2013.

\bibitem[Doblas-Reyes et~al.(2003)Doblas-Reyes, Pavan, and
  Stephenson]{doblas2003skill}
F.~Doblas-Reyes, V.~Pavan, and D.~Stephenson.
\newblock {The skill of multi-model seasonal forecasts of the wintertime North
  Atlantic Oscillation}.
\newblock \emph{Climate dynamics}, 21\penalty0 (5-6):\penalty0 501--514, 2003.

\bibitem[Eade et~al.(2014)Eade, Smith, Scaife, Wallace, Dunstone, Hermanson,
  and Robinson]{eade2014seasonal}
R.~Eade, D.~Smith, A.~Scaife, E.~Wallace, N.~Dunstone, L.~Hermanson, and
  N.~Robinson.
\newblock {Do seasonal-to-decadal climate predictions underestimate the
  predictability of the real world?}
\newblock \emph{Geophysical Research Letters}, 41\penalty0 (15):\penalty0
  5620--5628, 2014.

\bibitem[Efron and Tibshirani(1994)]{efron1994introduction}
B.~Efron and R.~J. Tibshirani.
\newblock \emph{{An Introduction to the Bootstrap}}.
\newblock CRC press, 1994.

\bibitem[Everitt(1984)]{everitt1984introduction}
B.~S. Everitt.
\newblock \emph{{An Introduction to Latent Variable Models}}.
\newblock Springer, 1984.

\bibitem[Feddersen et~al.(1999)Feddersen, Navarra, and
  Ward]{feddersen1999reduction}
H.~Feddersen, A.~Navarra, and M.~N. Ward.
\newblock Reduction of model systematic error by statistical correction for
  dynamical seasonal predictions.
\newblock \emph{J. Climate}, 12\penalty0 (7):\penalty0 1974--1989, Jul 1999.

\bibitem[Friedman et~al.(2009)Friedman, Hastie, and
  Tibshirani]{friedman2001elements}
J.~Friedman, T.~Hastie, and R.~Tibshirani.
\newblock \emph{{The Elements of Statistical Learning: Data Mining, Inference
  and Prediction}}, volume~2.
\newblock Springer, Berlin, 2009.
\newblock URL \url{http://statweb.stanford.edu/~tibs/ElemStatLearn/}.

\bibitem[Fuller(1987)]{fuller1987measurement}
W.~A. Fuller.
\newblock \emph{{Measurement Error Models}}.
\newblock John Wiley \& Sons, 1987.

\bibitem[Gelman and Robert(2013)]{gelman2013not}
A.~Gelman and C.~P. Robert.
\newblock {``Not Only Defended But Also Applied'': The Perceived Absurdity of
  Bayesian Inference}.
\newblock \emph{The American Statistician}, 67\penalty0 (1):\penalty0 1--5, Feb
  2013.

\bibitem[Gelman and Rubin(1992)]{gelman1992inference}
A.~Gelman and D.~B. Rubin.
\newblock {Inference from iterative simulation using multiple sequences}.
\newblock \emph{Statistical science}, pages 457--472, 1992.

\bibitem[Gelman et~al.(2004)Gelman, Carlin, Stern, and
  Rubin]{gelman2004bayesian}
A.~Gelman, J.~B. Carlin, H.~S. Stern, and D.~B. Rubin.
\newblock \emph{{Bayesian Data Analysis}}.
\newblock Chapman Hall/CRC, 2004.

\bibitem[Glahn and Lowry(1972)]{glahn1972mos}
H.~R. Glahn and D.~A. Lowry.
\newblock {The Use of Model Output Statistics (MOS) in Objective Weather
  Forecasting}.
\newblock \emph{Journal of Applied Meteorology}, 11\penalty0 (8):\penalty0
  1203--1211, Dec 1972.

\bibitem[Gneiting et~al.(2005)Gneiting, Raftery, Westveld~III, and
  Goldman]{gneiting2005calibrated}
T.~Gneiting, A.~E. Raftery, A.~H. Westveld~III, and T.~Goldman.
\newblock {Calibrated probabilistic forecasting using ensemble model output
  statistics and minimum CRPS estimation}.
\newblock \emph{Monthly Weather Review}, 133\penalty0 (5):\penalty0 1098--1118,
  2005.

\bibitem[Goddard et~al.(2013)Goddard, Kumar, Solomon, Smith, Boer, Gonzalez,
  Kharin, Merryfield, Deser, Mason, et~al.]{goddard2013verification}
L.~Goddard, A.~Kumar, A.~Solomon, D.~Smith, G.~Boer, P.~Gonzalez, V.~Kharin,
  W.~Merryfield, C.~Deser, S.~J. Mason, et~al.
\newblock {A verification framework for interannual-to-decadal predictions
  experiments}.
\newblock \emph{Climate Dynamics}, 40\penalty0 (1-2):\penalty0 245--272, 2013.

\bibitem[{Hurrell, James and National Center for Atmospheric Research Staff
  (Eds)}(2014)]{hurrell2014station}
{Hurrell, James and National Center for Atmospheric Research Staff (Eds)}.
\newblock {The Climate Data Guide: Hurrell North Atlantic Oscillation (NAO)
  Index (station-based)}.
\newblock Retrieved online from
  https://climatedataguide.ucar.edu/climate-data/hurrell-north-atlantic-oscillation-nao-index-station-based
  [last modified 05 Sep 2014], 2014.

\bibitem[Jaynes(2003)]{jaynes2003probability}
E.~T. Jaynes.
\newblock \emph{{Probability Theory: The Logic of Science}}.
\newblock Cambridge university press, 2003.

\bibitem[Jolliffe and Stephenson(2012)]{jolliffe2012forecast}
I.~T. Jolliffe and D.~B. Stephenson.
\newblock \emph{{Forecast Verification: A Practitioner's Guide in Atmospheric
  Science}}.
\newblock John Wiley \& Sons, 2012.

\bibitem[Kang et~al.(2014)Kang, Lee, Im, Kim, Kim, Kang, Schubert, Arribas, and
  MacLachlan]{kang2014prediction}
D.~Kang, M.-I. Lee, J.~Im, D.~Kim, H.-M. Kim, H.-S. Kang, S.~D. Schubert,
  A.~Arribas, and C.~MacLachlan.
\newblock Prediction of the arctic oscillation in boreal winter by dynamical
  seasonal forecasting systems.
\newblock \emph{Geophys. Res. Lett.}, 41\penalty0 (10):\penalty0 3577--3585,
  May 2014.

\bibitem[Kharin and Zwiers(2003)]{kharin2003improved}
V.~V. Kharin and F.~W. Zwiers.
\newblock {Improved seasonal probability forecasts}.
\newblock \emph{Journal of Climate}, 16\penalty0 (11):\penalty0 1684--1701,
  2003.

\bibitem[Kumar et~al.(2014)Kumar, Peng, and Chen]{kumar2014relationship}
A.~Kumar, P.~Peng, and M.~Chen.
\newblock {Is there a relationship between potential and actual skill?}
\newblock \emph{Monthly Weather Review}, 142\penalty0 (6):\penalty0 2220--2227,
  2014.

\bibitem[Lindley(2006)]{lindley2006understanding}
D.~V. Lindley.
\newblock \emph{Understanding uncertainty}.
\newblock John Wiley \& Sons, 2006.

\bibitem[MacLachlan et~al.(2014)MacLachlan, Arribas, Peterson, Maidens,
  Fereday, Scaife, Gordon, Vellinga, Williams, Comer, and
  et~al.]{maclachlan2014global}
C.~MacLachlan, A.~Arribas, K.~A. Peterson, A.~Maidens, D.~Fereday, A.~A.
  Scaife, M.~Gordon, M.~Vellinga, A.~Williams, R.~E. Comer, and et~al.
\newblock {Global Seasonal forecast system version 5 (GloSea5): A
  high-resolution seasonal forecast system}.
\newblock \emph{Quarterly Journal of the Royal Meteorological Society}, Jun
  2014.

\bibitem[Madden(1976)]{madden1976estimates}
R.~A. Madden.
\newblock {Estimates of the natural variability of time-averaged sea-level
  pressure}.
\newblock \emph{Monthly Weather Review}, 104\penalty0 (7):\penalty0 942--952,
  1976.

\bibitem[Mardia et~al.(1979)Mardia, Kent, and Bibby]{mardia1979multivariate}
K.~V. Mardia, J.~T. Kent, and J.~M. Bibby.
\newblock \emph{{Multivariate Analysis}}.
\newblock Academic press, 1979.

\bibitem[Moran(1971)]{moran1971estimating}
P.~Moran.
\newblock {Estimating structural and functional relationships}.
\newblock \emph{Journal of Multivariate Analysis}, 1\penalty0 (2):\penalty0
  232--255, 1971.

\bibitem[Murphy and Wilks(1998)]{murphy1998case}
A.~H. Murphy and D.~S. Wilks.
\newblock A case study of the use of statistical models in forecast
  verification: Precipitation probability forecasts.
\newblock \emph{Weather and Forecasting}, 13\penalty0 (3):\penalty0 795--810,
  1998.

\bibitem[Murphy and Winkler(1987)]{murphy1987general}
A.~H. Murphy and R.~L. Winkler.
\newblock {A general framework for forecast verification}.
\newblock \emph{Monthly Weather Review}, 115\penalty0 (7):\penalty0 1330--1338,
  1987.

\bibitem[Murphy(1990)]{murphy1990assessment}
J.~M. Murphy.
\newblock Assessment of the practical utility of extended range ensemble
  forecasts.
\newblock \emph{Quarterly Journal of the Royal Meteorological Society},
  116\penalty0 (491):\penalty0 89--125, Jan 1990.
\newblock ISSN 1477-870X.
\newblock \doi{10.1002/qj.49711649105}.
\newblock URL \url{http://dx.doi.org/10.1002/qj.49711649105}.

\bibitem[{National Center for Atmospheric Research Staff
  (Eds)}(2014)]{ncar2014pc}
{National Center for Atmospheric Research Staff (Eds)}.
\newblock {The Climate Data Guide: Hurrell North Atlantic Oscillation (NAO)
  Index (PC-based)}.
\newblock Retrieved online from
  https://climatedataguide.ucar.edu/climate-data/hurrell-north-atlantic-oscillation-nao-index-pc-based
  [last modified 05 Sep 2014], 2014.

\bibitem[Otto et~al.(2012)Otto, Ferro, Fricker, and Suckling]{otto2012judging}
F.~E. Otto, C.~Ferro, T.~Fricker, and E.~Suckling.
\newblock {On judging the credibility of climate projections}.
\newblock \emph{Climatic Change}, 2012.

\bibitem[Pearl(2000)]{pearl2000causality}
J.~Pearl.
\newblock \emph{{Causality: Models, Reasoning and Inference}}.
\newblock Cambridge Univ Press, 2000.

\bibitem[Riddle et~al.(2013)Riddle, Butler, Furtado, Cohen, and
  Kumar]{riddle2013cfsv2}
E.~E. Riddle, A.~H. Butler, J.~C. Furtado, J.~L. Cohen, and A.~Kumar.
\newblock Cfsv2 ensemble prediction of the wintertime arctic oscillation.
\newblock \emph{Climate Dynamics}, 41\penalty0 (3-4):\penalty0 1099--1116, Jul
  2013.

\bibitem[Robert(2007)]{robert2007bayesian}
C.~P. Robert.
\newblock \emph{{The Bayesian Choice: From Decision-Theoretic Foundations to
  Computational Implementation (Springer Texts in Statistics)}}.
\newblock Springer New York, 2007.

\bibitem[Rougier et~al.(2013)Rougier, Goldstein, and House]{rougier2013second}
J.~Rougier, M.~Goldstein, and L.~House.
\newblock {Second-Order Exchangeability Analysis for Multimodel Ensembles}.
\newblock \emph{Journal of the American Statistical Association}, 108\penalty0
  (503):\penalty0 852--863, 2013.

\bibitem[Roulston and Smith(2002)]{roulston2002evaluating}
M.~S. Roulston and L.~A. Smith.
\newblock {Evaluating probabilistic forecasts using information theory.}
\newblock \emph{Monthly Weather Review}, 130\penalty0 (6), 2002.

\bibitem[Sansom et~al.(2013)Sansom, Stephenson, Ferro, Zappa, and
  Shaffrey]{sansom2013simple}
P.~G. Sansom, D.~B. Stephenson, C.~A. Ferro, G.~Zappa, and L.~Shaffrey.
\newblock {Simple uncertainty frameworks for selecting weighting schemes and
  interpreting multimodel ensemble climate change experiments}.
\newblock \emph{Journal of Climate}, 26\penalty0 (12):\penalty0 4017--4037,
  2013.

\bibitem[Scaife et~al.(2014)Scaife, Arribas, Blockley, Brookshaw, Clark,
  Dunstone, Eade, Fereday, Folland, Gordon, et~al.]{scaife2014skillful}
A.~Scaife, A.~Arribas, E.~Blockley, A.~Brookshaw, R.~Clark, N.~Dunstone,
  R.~Eade, D.~Fereday, C.~Folland, M.~Gordon, et~al.
\newblock {Skillful long-range prediction of European and North American
  winters}.
\newblock \emph{Geophysical Research Letters}, 41\penalty0 (7):\penalty0
  2514--2519, 2014.

\bibitem[Shi et~al.(2015)Shi, Schaller, MacLeod, Palmer, and
  Weisheimer]{shi2015impact}
W.~Shi, N.~Schaller, D.~MacLeod, T.~Palmer, and A.~Weisheimer.
\newblock {Impact of hindcast length on estimates of seasonal climate
  predictability}.
\newblock \emph{Geophys. Res. Lett.}, Feb 2015.

\bibitem[Smith et~al.(2014)Smith, Scaife, Eade, and Knight]{smith2014seasonal}
D.~M. Smith, A.~A. Scaife, R.~Eade, and J.~R. Knight.
\newblock {Seasonal to decadal prediction of the winter North Atlantic
  Oscillation: Emerging capability and future prospects}.
\newblock \emph{Quarterly Journal of the Royal Meteorological Society}, 2014.

\bibitem[{Stan Development Team}(2014{\natexlab{a}})]{rstan2014}
{Stan Development Team}.
\newblock {RStan: the R interface to Stan, Version 2.5.0}, 2014{\natexlab{a}}.
\newblock URL \url{http://mc-stan.org/rstan.html}.

\bibitem[{Stan Development Team}(2014{\natexlab{b}})]{stan2014}
{Stan Development Team}.
\newblock \emph{{Stan Modeling Language Users Guide and Reference Manual,
  Version 2.5.0}}, 2014{\natexlab{b}}.
\newblock URL \url{http://mc-stan.org/}.

\bibitem[Stephenson et~al.(2012)Stephenson, Collins, Rougier, and
  Chandler]{stephenson2012statistical}
D.~B. Stephenson, M.~Collins, J.~C. Rougier, and R.~E. Chandler.
\newblock {Statistical problems in the probabilistic prediction of climate
  change}.
\newblock \emph{Environmetrics}, 23\penalty0 (5):\penalty0 364--372, 2012.

\bibitem[Tebaldi et~al.(2005)Tebaldi, Smith, Nychka, and
  Mearns]{tebaldi2005quantifying}
C.~Tebaldi, R.~L. Smith, D.~Nychka, and L.~O. Mearns.
\newblock {Quantifying uncertainty in projections of regional climate change: A
  Bayesian approach to the analysis of multimodel ensembles}.
\newblock \emph{Journal of Climate}, 18\penalty0 (10):\penalty0 1524--1540,
  2005.

\bibitem[Tippett et~al.(2007)Tippett, Barnston, and
  Robertson]{tippett2007estimation}
M.~K. Tippett, A.~G. Barnston, and A.~W. Robertson.
\newblock Estimation of seasonal precipitation tercile-based categorical
  probabilities from ensembles.
\newblock \emph{J. Climate}, 20\penalty0 (10):\penalty0 2210--2228, May 2007.

\bibitem[Von~Storch and Zwiers(2001)]{storch2001statistical}
H.~Von~Storch and F.~W. Zwiers.
\newblock \emph{{Statistical analysis in climate research}}.
\newblock {Cambridge University Press}, 2001.

\bibitem[Wang and Bishop(2005)]{wang2005improvement}
X.~Wang and C.~H. Bishop.
\newblock Improvement of ensemble reliability with a new dressing kernel.
\newblock \emph{Quarterly Journal of the Royal Meteorological Society},
  131\penalty0 (607):\penalty0 965--986, 2005.

\bibitem[Weigel et~al.(2009)Weigel, Liniger, and
  Appenzeller]{weigel2009seasonal}
A.~P. Weigel, M.~A. Liniger, and C.~Appenzeller.
\newblock {Seasonal ensemble forecasts: Are recalibrated single models better
  than multimodels?}
\newblock \emph{Monthly Weather Review}, 137\penalty0 (4):\penalty0 1460--1479,
  2009.

\end{thebibliography}

\end{document}